\RequirePackage{fix-cm}
\documentclass[smallextended]{svjour3}       
\smartqed  
\usepackage{graphicx}
\usepackage[font=small,labelfont=bf]{caption}
\usepackage{tabularx}
\usepackage{ragged2e}
\newcolumntype{Y}{>{\RaggedLeft}X}
\usepackage{booktabs}
\usepackage{pdflscape}
\usepackage{amsmath}
\usepackage[bbgreekl]{mathbbol}

\usepackage{hyperref}
\hypersetup{
  colorlinks   = true, 
  urlcolor     = blue, 
  linkcolor    = blue, 
  citecolor   = red 
}

\journalname{Quantum Science and Technology}
\begin{document}

\title{For high-precision bosonic Josephson junctions, many-body effects matter 
}


\author{Marie A. McLain        \and
        Diego A. Alcala        \and
        Lincoln D. Carr 
}


\institute{Marie A. McLain \and Diego A. Alcala \and Lincoln D. Carr \at
              Department of Physics, Colorado School of Mines, Golden, Colorado, USA\\
}

\date{Received: date / Accepted: date}

\maketitle

\begin{abstract}
Typical treatments of superconducting or superfluid Josephson junctions rely on mean-field or two-mode models; we explore many-body dynamics of an isolated, ultracold, Bose-gas long Josephson junction using time-evolving block decimation simulations. We demonstrate that with increasing repulsive interaction strength, localized dynamics emerge that influence macroscopic condensate behavior and can lead to formation of solitons that directly oppose the symmetry of the junction. Initial state population and phase yield insight into dynamic tunneling regimes of a quasi one-dimensional double well potential, from Josephson oscillations to macroscopic self-trapping. Population imbalance simulations reveal substantial deviation of many-body dynamics from mean-field Gross-Pitaevskii predictions, particularly as the barrier height and interaction strength increase. In addition, the sudden approximation supports localized particle-hole formation after a diabatic quench, and correlation measures unveil a new dynamic regime: the Fock flashlight.

\end{abstract}

\section{Introduction}
\label{sec:intro}
Ground states of linear systems comply with the symmetry of their underlying confining potential. For example, in a one-dimensional double well, the ground state is symmetric about the barrier and the orthogonal state is antisymmetric. Spontaneous symmetry breaking in such a potential means that tuning a control parameter beyond a critical value leaves the system with two concurrent ground states that no longer conform to the symmetry of the double well: each ground state is asymmetric. In an isolated Josephson junction such as that realized in Bose-Einstein condensates (BECs), spontaneous symmetry breaking can be observed as a $\mathbb{Z}_2$ quantum phase transition (QPT) from Josephson oscillations to a macroscopic self-trapping phase \cite{albiez2005,malomed2014}. The effect is observed in many physical systems, particularly nonlinear optics \cite{du2016,malomed2015} and BECs \cite{malomed2014,gaury2015} where recent experimental progress has accelerated \cite{baumann2010,alberucci2015,trenkwalder2016}. For example, this symmetry breaking offers coexistent states that can be used in quantum memory applications such as quantum flip-flops \cite{hamel2015,malomed2015}. In superconducting systems, the Coulomb blockade effect occurs when the critical parameter exceeds that of the external current bias and holds exciting prospects for use as a noise filter \cite{sonin2004,Haviland1991,Geerligs1989,Haviland2000}. Superconducting Josephson junctions can be compared for example to driven bosonic Josephson junctions, where the population imbalance between the left and right wells differs as a result of this external bias \cite{Khomeriki2007}; in this paper we investigate instead an isolated bosonic Josephson junction. In order for full characterization of such devices, the many-body influences in dynamic regimes must first be understood, and BECs provide a promising architecture to do so as precise quantum simulators.

BECs pose a highly-controllable mechanism for probing the many-body effects that become crucial in high-precision applications: interactions can be swept over seven orders of magnitude with Feshbach resonance manipulation of scattering length \cite{inouye1998,feshbach}. The advancement of experimental technique in radio frequency magnetic traps \cite{magneticTrap} together with improvements in optical traps \cite{opticalTrap} enable enhanced control of experiments and provide systems devoid of defects, which is critical for highly precise characterization of dynamics. In addition, Bose-Einstein condensates in optical lattices facilitate the measuring and manipulating of many body quantum states \cite{manyBodyExample,bakr2009,sherson2010} that remain intractable in many other experimental platforms. Therefore, BECs such as those formed by $^{87}$Rb are an ideal backdrop for investigating the dynamics of a long bosonic Josephson junction, or a double well with spatial extent. The spatial extent of the double well is experimentally imposed in a 1D waveguide with an optical lattice, where the transverse degrees of freedom have been suppressed \cite{Olshanii1998}. This underlying lattice enables precise study of the many-body interactions. We represent the discrete nature of the lattice by the Bose-Hubbard Hamiltonian (BHH),

\begin{equation}
\label{eqn:BHH}
\hat{H}_{\mathrm{BH}} = {} -J\sum_{\langle i,j\rangle}(\hat{b}^{\dagger}_i \hat{b}_{j}+\hat{b}_i \hat{b}^{\dagger}_{j}) +
\frac{1}{2}U\sum_i \hat{n}_i(\hat{n}_i-1) + 
\sum_i V_i\hat{n}_i\,,
\end{equation}
where $J$ is the bosonic tunneling strength, $U$ is the interaction strength, $\hat{b}^{\dagger}_i$ and $\hat{b}_i$ are bosonic creation and destruction operators, respectively, satisfying bosonic commutation relations, $\hat{n}_i$ is the bosonic number operator, $\langle i,j\rangle$ indicates nearest neighbors, and the indices $i,j \epsilon \{1,...,L\}$ run over the 1D lattice of length $L$.  $V_i$ is the height of the external double well potential. While experimental double wells may be implemented with smoother potentials, by describing the barrier with a single parameter, the barrier height, we minimize the size of the parameter space and thus increase numerical efficiency. It is important to note that the shape of the potential may influence the dynamics \cite{Khomeriki2007}, and in our case the barrier is square and thin compared to the length of the junction, which relates to e.g. the weak link in superconducting Josephson junctions \cite{nistjj}. This BHH model is applicable when the lattice sites are sufficiently deep to allow for tight binding and single band approximations \cite{jaksch1998}.

For unit filling with fixed particle number, the optical lattice introduces a second quantum phase transition (QPT) to the Josephson junction in addition to the well-known $\mathbb{Z}_2$ transition, a continuous $U(1)$ transition from a Mott insulator to a superfluid \cite{kuhner1998,lacki2016}. In the BHH for $V_i=0$, $L \to \infty$, $N \to \infty$, $N/L = 1$, where $N$ is the number of atoms, the crossover from the superfluid phase to the Mott insulating phase occurs at a $J/U_{\mathrm{critical}} \approx 0.305 $ in one dimension \cite{Cristiani2002}, while the mean-field approximation underestimates this at $J/U_{\mathrm{critical}} \approx 0.086 $ \cite{fisher1989}. Superfluid behavior manifests for weakly-interacting systems such that $J/U > J/U_{\mathrm{critical}}$; in contrast, the Mott-insulating regime requires strongly-interacting bosons such that $J/U < J/U_{\mathrm{critical}}$. When the lattice is infinitely deep, or $J=0$, the result is a perfect Fock state; in Fock space notation this state is written $\mid 1 1 1...1 \rangle$ for unit filling, i.e., the number of particles is approximately commensurate with the number of lattice sites. In this paper we focus on 1D optical lattices, which translates to experiment e.g. via ``cigar''-shaped waveguides, where the atoms are contained in one-dimensional potentials by transverse optical confinement \cite{Olshanii1998}.

While it is convenient for our research to observe bosons in a Fock or number basis, often it is more appropriate to use a different framework or basis, for example in cases of macroscopic phase coherence and Josephson-like dynamics, the phase basis is often represented as a phase difference between the two wells of a double well potential. The dictionary in Table \ref{table:hamstable} gives a flavor of common models of Josephson junctions and their relation to one another. The first Hamiltonian is the Bose-Hubbard model; it differs from the other Hamiltonians in that it is presented as a discretization scheme, where tunneling strength $J$ and atomic interaction strength $U$ are local and are defined by integrals over lowest-band Wannier functions centered on periodic lattice sites, $w_0(\vec{x})$; alternately, as in our present study, we may take the lattice as an explicitly imposed potential. The second Hamiltonian, the two-mode model presented in Table \ref{table:hamstable}(2) is the same as the Bose-Hubbard Hamiltonian, but with only two lattice sites, one for each well. The general phase and number field operators are conjugate variables, such that $[\hat{\phi},\hat{N}]=-i$, and the raising and lowering operators to obtain Table \ref{table:hamstable}(3) and (4) are $e^{i\hat{\phi}}$ and $e^{-i\hat{\phi}}$, respectively. In the general double well problem, mean-field theory provides a description of the relative dynamics under certain assumptions: the coherence must be large and relative phase well-defined. This mean-field approximation is generally very good in regimes where interactions are weak, such as the superfluid or superconducting regimes \cite{sonin2004,tinkham2004,Haviland2000,Levy2007}.

\thispagestyle{empty}
\begin{table}
\caption[Dictionary of analogous Josephson junction naming conventions.]{\label{table:hamstable} \textit{Translation dictionary of Josephson junction naming conventions.} (1) The Bose-Hubbard Hamiltonian used to describe Fock space number operators on an optical lattice used as a discretization scheme. (2) The Lipkin-Meshkov-Glick model is a two-mode model, which as a distinction from the BHH does not capture many-body effects in a long BJJ; it refers to two macroscopic Wannier functions in a Fock basis, where the tunneling parameter is the Josephson or double well tunneling energy $E_{\mathrm{DWT}}$, and the the inter-atomic charging energy, $\tilde{E}_C$, is between particles in the same well. (3) The many-body relative number, relative phase formulation in a phase basis allows for number fluctuations. (4) The mean-field limit of (3), when the number of particles becomes large, is a semiclassical description void of number fluctuations. (5) The Hamiltonian for an unbiased Josephson junction in a superconducting circuit or an atomic gas with an applied current takes an analogous form as (4), where the change in representation here is the charging energy, $1/[2C]$, which is the capacitance between Cooper pairs on opposing sides of the junction. Finally, (6) a phase qubit Hamiltonian includes the same tunneling and interaction terms as the mean-field Josephson junction, with the benefit of an added current bias term that tunes the tilt of the potential in the phase basis.}
\resizebox{1.0\linewidth}{!}{
\begin{tabularx}{1.08\linewidth}[h]{@{} Y@{\hskip 15pt} Y@{\hskip 14pt} Y@{\hskip 8pt} c@{}}
\addlinespace
\toprule
\textbf{Double Well Convention} & \textbf{Hamiltonian} & \textbf{Tunneling Parameter} & \textbf{Interaction Parameter} \\
\midrule
(1) Bose Hubbard as Discretization & $\hat{H}_{\mathrm{BH}} = -J\sum_i(\hat{b}^{\dagger}_i \hat{b}_{i+1}+\mathrm{h.c.}) + \frac{1}{2}U\sum_i \hat{n}_i(\hat{n}_i-1) + \sum_i V_i\hat{n}_i$ & $J = -\int d\vec{x} w_0^{\ast}(\vec{x}-\vec{x}_i)(-\frac{\hbar^2}{2m}\vec{\nabla}^2+V_{\mathrm{latt}}(\vec{x}))w_0(\vec{x}-\vec{x}_{i+1})$ & $U = g \int d\vec{x}  \mid{w_0(\vec{x}_i)}\mid^4$ \\ \addlinespace \addlinespace
(2) Lipkin-Meshkov-Glick & $\hat{H}_{\mathrm{LMG}} = -E_{\mathrm{DWT}} (\hat{b}^{\dagger}_L \hat{b}_{R}+\hat{b}_L \hat{b}^{\dagger}_{R}) + \frac{1}{2} \tilde{E}_C (\hat{n}_L(\hat{n}_L-1)+\hat{n}_R(\hat{n}_R-1))$ & $N E_{\mathrm{DWT}}$ & $\tilde{E}_C$ \\ \addlinespace \addlinespace
(3) Relative Number/Phase & $\hat{H}_{\hat{n}\hat{\phi}} = -E_{\mathrm{DWT}} \sqrt{N(N+2)-4\langle\hat{n}^2\rangle} \cos \hat{\phi}  + \frac{1}{2}\tilde{E}_C \hat{n}(\hat{n}-1)$ & $E_{\mathrm{DWT}}$ & $\tilde{E}_C$ \\ \addlinespace \addlinespace
(4) Semiclassical Rel. Number/Phase & $H_{\mathrm{iso}} = -E_{\mathrm{DWT}} \sqrt{1-n^2} \cos \phi +
\frac{\tilde{E}_C}{2} n^2$ & $E_{\mathrm{DWT}}$ & $\tilde{E}_C$ \\ \addlinespace \addlinespace
(5) Unbiased Superconducting JJ & $\hat{H}_{\mathrm{JJ}} = -\frac{I_c \Phi_0}{2\pi}\cos\hat{\phi} + \frac{1}{2C}\hat{Q}^2$ & $I_c \Phi_0/2\pi$ & $\frac{1}{4e^2C}$ \\ \addlinespace \addlinespace
(6) Superconducting Phase Qubit & $\hat{H}_{\mathrm{PQ}} = -\frac{I_c \Phi_0}{2\pi}\cos\hat{\phi} + \frac{1}{2C}\hat{Q}^2-\frac{I_{\mathrm{bias}}\Phi_0}{2\pi}\hat{\phi}$ & $I_c \Phi_0/2\pi$ & $\frac{1}{4e^2C}$ \\
\bottomrule
\end{tabularx}}

\end{table} 

The tunneling dynamics are underlaid by the interplay of the two QPTs, $\mathbb{Z}_2$ and U(1). This reduces to tuning of the barrier height for the spontaneous symmetry breaking transition and tuning of the interaction strength for the superfluid-Mott transition. The ultimate goal then is to characterize the dynamic tunneling regimes that result, from Josephson oscillations to self-trapping. Some of these regimes have previously been identified in literature for weakly-interacting systems \cite{albiez2005,Levy2007,Galbiati2012,gaury2015,malomed2014}. In a single-particle limit, the Rabi frequency is dependent on $E_{\mathrm{DWT}}$, the Josephson tunneling energy, $\omega_R \propto 2 E_{\mathrm{DWT}}/ L \hbar$, where the frequency decreases with increasing system size and $L$ is the number of lattice sites \cite{abbarchi2013,yu2002}. 

The mean-field equations of motion, from \cite{Raghavan1999}, are similar to Josephson's equations; the major difference stems from the assumption in Josephson's equations that the time derivative of population density is identically the same in the two wells \cite{Smerzi1997,Raghavan1999}. Relinquishing this restriction elicits the equations of motion:

\begin{equation}
\begin{aligned}
n'(t) & {} = -\sqrt{(1-n(t)^2)}\sin{\phi(t)} \\
\phi'(t) & = \Delta E + E_C n(t) + \frac{n(t)}{\sqrt{1-n(t)^2}} \cos{\phi(t)}.
\end{aligned}
\end{equation}
The true Rabi or sinusoidal regime occurs when the charging energy is zero, $\tilde{E}_C = 0$, since the charging energy term in the mean-field Hamiltonian (unbiased JJ (4 and 5) in the Table \ref{table:hamstable}) provides a nonlinear effect that takes us out of the Rabi and into the Josephson regime \cite{Raghavan1999,albiez2005}. Josephson oscillations have a frequency proportional to $\omega_J =\sqrt{\tilde{E}_C E_{\mathrm{DWT}}}/\hbar$, where $\tilde{E}_C$ is an inter-atomic charging energy between bosonic atoms. A mean-field measure often used to characterize Josephson dynamics is the relative particle number between the two wells, or the population imbalance, $(n_L(t) - n_R(t))/N_{\mathrm{tot}}$, a variable that is close to zero in superconducting circuits due to a dominating external current \cite{feynman1965}. The initial population imbalance, $n_0$, and relative overall phase between the two wells, $\phi_0$, provide information to predict the dynamic tunneling regime. When $n_0$ exceeds a critical threshold, the particles will be self-trapped on one side of the junction, thus breaking the ground state symmetry in a $\mathbb{Z}_2$ transition. This out-of-equilibrium phenomenon renormalizes the energy to a metastable tunneling-suppressed Fock regime \cite{abbarchi2013}.

To characterize these dynamic regimes in relation to the superfluid-Mott phase transition, we annotate the relevant parameter space. First, we have the BHH describing the underlying optical lattice, with parameters $J/U$, where $J$ is the local bosonic tunneling or hopping strength between lattice sites and $U$ is the local bosonic interaction strength between atoms on the same lattice site. Next, we have a two-mode model with $E_{\mathrm{DWT}}$  the tunneling between wells, or Josephson energy, and $\tilde{E}_C$ the interaction between atoms in the same well, a modified charging energy. $E_{\mathrm{DWT}}/\tilde{E}_C$ parameters together represent the ratio of tunneling between the two wells and interaction within each well.  $E_{\mathrm{DWT}}/\tilde{E}_C$ reduces to a function of $\zeta$, the effective energy ratio for the two wells (see Equations (\ref{eqn:zMF}), (\ref{eqn:ztot}), (\ref{eqn:zMB}) below), when the width of the barrier is held constant. The nature of the two scales of the problem, in addition to system size, leads to multiscale behavior.

\begin{figure}
\centering
\includegraphics[width=0.91\textwidth]{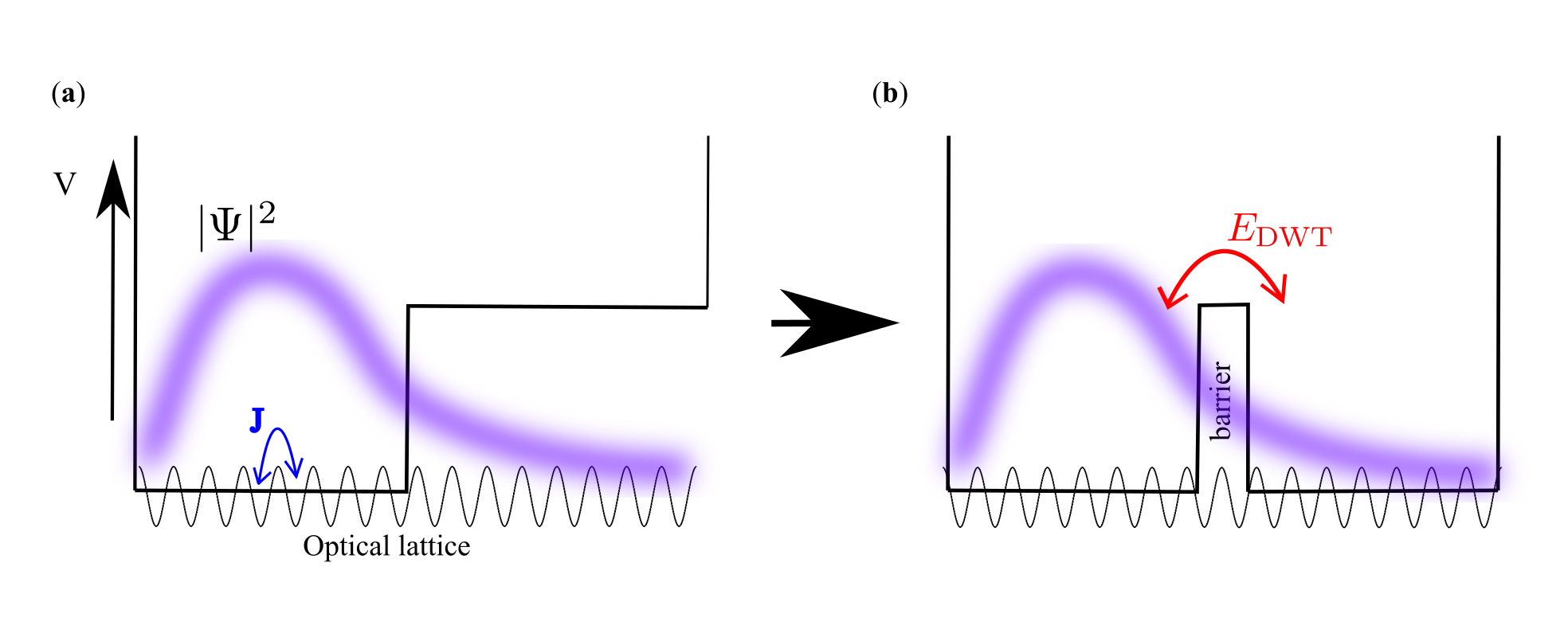}
\caption[Initial bound state and time propagation.]{\label{fig:ITPRTP}\textit{Initial bound state and time propagation.} (\textbf{a}) We run imaginary time propagation in the potential (solid black curve) to obtain a ground state wavefunction (blurred purple curve) of the left well, (\textbf{b}) then lower the right well and propagate in real time. The optical lattice is superimposed on the Josephson junction as a discretization scheme.}
\end{figure}

To initialize a state for the dynamics, we run imaginary time propagation to obtain a ground state of a single well potential as shown in Figure \ref{fig:ITPRTP}(\textbf{a}). The finite potential on the right in this instance is at the same height as the double well barrier. Then, we diabatically quench to a symmetric bosonic Josephson junction by lowering the right side potential before propagating in real time, as portrayed in Figure \ref{fig:ITPRTP}(\textbf{b}). This protocol is performed both in time-evolving block decimation (TEBD) and in solving the Gross-Pitaevskii equation (GPE) as a mean-field approximation for symmetric double wells with a single lattice site as the barrier and with open boundary conditions. The system sizes range from four sites for analytical sudden approximation calculations to 55 sites for $g^{(2)}$ number fluctuations calculated with TEBD.

This paper is outlined as follows. Results are presented first in terms of static initial states in section \ref{sec:bjjinitialsts}. We propose a method for characterizing the dynamic regimes based on the initial state number and phase from both a mean-field and a many-body perspective, and we demonstrate that the mean-field representation fails to recognize the self-trapping transition above a weak-interaction threshold. For strongly-interacting bosons, the diabatic quench instigates particle-hole creation and low-lying junction modes; the result is corroborated by sudden approximation calculations. In section \ref{sec:dynamicFock}, we find the particle-hole pairs, which are number-squeezed, are exceptionally stable over long time scales and are particularly pronounced in the Josephson regime. When the filling factor is slightly above or below one, we observe what appear to be emergent solitons or solitary waves -- they form as pairs on either side of the barrier and propagate in parallel without dispersing and with a speed less than the sound speed -- a striking contrast to the symmetry of the underlying potential. Furthermore, we compare the dynamical similarities of TEBD and the GPE for weakly-interacting particles, along with the failure of the GPE to capture the $\mathbb{Z}_2$ transition for strongly-interacting bosons. Section \ref{sec:dynamicCorrelations} focuses on correlation measures, which are easily calculated from matrix product states and illuminate the many-body physics especially as interaction strengths increase. For example, the depletion of the condensate as a function of barrier height becomes quantized in the Mott regime due to the Mott gap, whereas in the superfluid regime the BEC is universally strong except near the symmetry-breaking critical point. We also present the time-dependence of the $g^{(2)}$ measure, or number fluctuations, through snapshots that can be directly compared with experimental results. Then, in the materials and methods of section \ref{sec:matmethods}, we present more detail on our open-source time-evolving block decimation simulations and segue to open-source matrix product state (OSMPS) software. We also describe our protocol for numerically solving the GPE and our analytical sudden approximation calculations, as well as how to access our open data repository. In the final section \ref{sec:conclusion}, we conclude with a brief summary of our results, apply our findings in a broader context, and suggest future research direction.

\section{Results and Discussion}
\label{sec:results}

\subsection{Initial state influence on dynamics}
\label{sec:bjjinitialsts}

While typical treatments of the dynamics of quantum phase transitions focus on a quench of the parameter space using ground states, the goal of this work is to characterize the dynamical parameter space associated with macroscopic quantum tunneling.  Using the initial states, we can make many predictions about the dynamics, though behavior such as soliton creation or number fluctuation propagation do not follow from from initial states alone.

We devise a method that maps out dynamical regimes, both from a many-body and a mean-field perspective, based on initial state number and phase information. To quantify the competing energy terms, we define a mean-field and many-body version of a critical parameter, which is a ratio of tunneling and interaction energies, to distinguish between Josephson and self-trapping or Fock regimes. When the energy ratio exceeds one, $\zeta > 1$, the system exhibits spontaneous symmetry breaking and the particles remain confined on one side of the potential. The mean-field version is based on an energy ratio suggested in \cite{Smerzi1997}. For initial population imbalance $n_0$ and initial relative phase $\phi_0$, we define the mean-field (MF) energy ratio from Hamiltonian (4) of Table \ref{table:hamstable},

\begin{equation}
\zeta_{\mathrm{MF}} = \frac{n_0^2}{2(1 + \sqrt{1 - n_0^2} \cos \phi_0)}
\label{eqn:zMF}
\end{equation}
where the initial state has a well-defined relative number and phase. 

From Hamiltonian (1) in Table \ref{table:hamstable}, we calculate a many-body (MB) version of Equation (\ref{eqn:zMF}), where we have not yet subtracted to obtain the relative phase and number information,

\begin{equation}
\begin{aligned}
\zeta_{\mathrm{tot}} & {} =  \frac{\frac{1}{2} \sum_j n_j^{(0)} (n_j^{(0)}-1)}{\sum_{\langle j,k \rangle} A_{jk} (\exp[i \phi_{jk}^{(0)}] + \exp[-i \phi_{jk}^{(0)}])} \\
& = \frac{\frac{1}{2} \sum_j n_j^{(0)} (n_j^{(0)}-1)}{\sum_{\langle j,k \rangle} 2 A_{jk} \cos{\phi_{jk}^{(0)}}}\\
\end{aligned}
\label{eqn:ztot}
\end{equation}
where $n_j^{(0)} \equiv \langle \hat{n}_j (t=0) \rangle$ is the initial number on site $j$ and $\phi_{jk}^{(0)}$ is the initial phase between nearest neighbor sites $j$ and $k$. The superscript $^{(0)}$ is a label indicating an initial value. The denominator of Equation (\ref{eqn:ztot}) is calculated from the single particle density matrix (SPDM), $\rho_{jk} = \langle \hat{b}^{\dagger}_j \hat{b}_k \rangle = A_{jk} \exp(i \phi_{jk})$ between nearest neighbor sites $j$ and $k$,  which comprises the tunneling term of the BHH (see Table \ref{table:hamstable}) without the tunneling constant $J$. The phase is calculated as follows. We take the eigenvector $\vec{v}^{(1)}$ associated with the largest eigenvalue $\lambda_1$ of the SPDM, where the vector indicates distribution over lattices sites of the form $v^{(1)}_j$, $j\in\{1,\ldots,L\}$.  This eigenvector is the closest approximation to the mean-field in the Landau definition of the superfluid order parameter, discretized on the lattice as $v^{(1)}_j = \psi_j = \sqrt{n_j}\exp{i\phi_j}$.  The phase is thus taken as the phase of the dominant eigenmode of the SPDM, and the relative phase is calculated as the difference between the average phase in the left and right wells, $\Delta\phi = L^{-1}_{\mathrm{left well}}\sum_{j\in\mathrm{left well}}\phi_j - L^{-1}_{\mathrm{right well}}\sum_{j\in\mathrm{right well}}\phi_j$ with $L_{\mathrm{left well}}$ the number of sites in the left well, and likewise for the right well.  Alternately one can calculate a particular phase difference, e.g. between the sites in the left and right wells closest to the barrier, respectively. The cosine representation of the tunneling term is a reminder of the analogies presented in Table \ref{table:hamstable}. Then, the numerator of Equation (\ref{eqn:ztot}) comprises the interaction term of the BHH without the constant $U$. Next, we define the initial SPDM as $\rho_{jk}^{(0)} \equiv \langle \hat{b}^{\dagger}_j (0) \hat{b}_k (0) \rangle $.  This leads to a many-body ratio analogous to the mean-field ratio of Equation (\ref{eqn:zMF}), where we calculate the difference of the interaction terms (tunneling terms) of the BHH between the two wells in the numerator (denominator) of Equation (\ref{eqn:zMB}). We express $\zeta_{\mathrm{MB}}$ as these many-body differences between the two wells,

\begin{equation}
\begin{aligned}
\zeta_{\mathrm{MB}} & {} =  \frac{\sum_{j=1}^{\left \lfloor{L/2}\right \rfloor} n_j^{(0)} (n_j^{(0)}-1) - \sum_{j=\left \lfloor{L/2}\right \rfloor}^{L} n_j^{(0)} (n_j^{(0)}-1)}{4 \biggl( \sum_{\langle j,k \rangle =1}^{\left \lfloor{L/2}\right \rfloor} (\rho_{jk}^{(0)}+\rho_{jk}^{(0)*}) - \sum_{\langle j,k \rangle = \left \lfloor{L/2}\right \rfloor}^{L} ( \rho_{jk}^{(0)}+\rho_{jk}^{(0)*}) \biggr) }\\
\end{aligned}
\label{eqn:zMB}
\end{equation}
where $\rho_{jk}^{(0)}$ is a complex element of the single particle density matrix for lattice sites $j$ and $k$ at time $t=0$ and $L$ is the total number of lattice sites. The sums $\sum_{j=1}^{\left \lfloor{L/2}\right \rfloor}$ and $\sum_{\langle j,k \rangle =1}^{\left \lfloor{L/2}\right \rfloor}$ are sums over the left well, and $\sum_{j=\left \lfloor{L/2}\right \rfloor}^{L}$ and $\sum_{\langle j,k \rangle = \left \lfloor{L/2}\right \rfloor}^{L}$ are sums over the right well.

\begin{figure}
\centering
\includegraphics[width=0.46\textwidth]{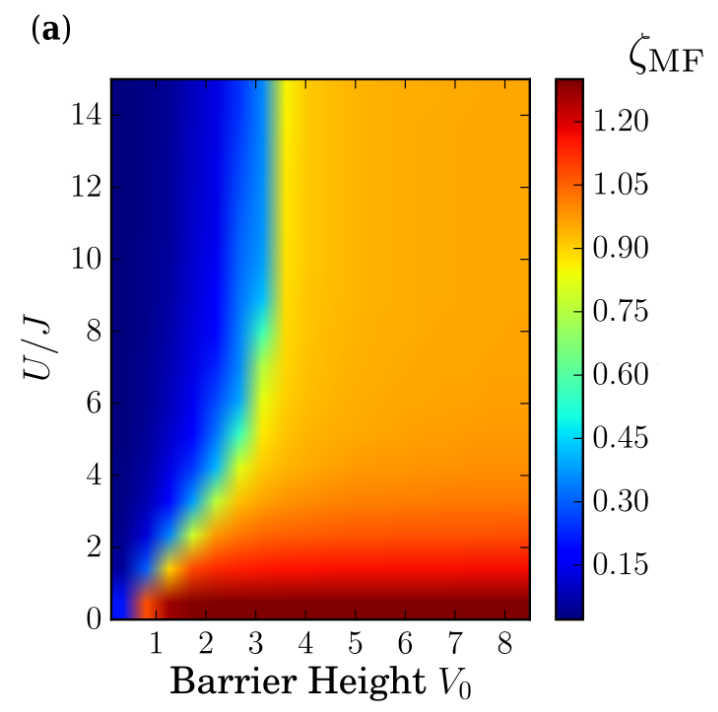}
\includegraphics[width=0.45\textwidth]{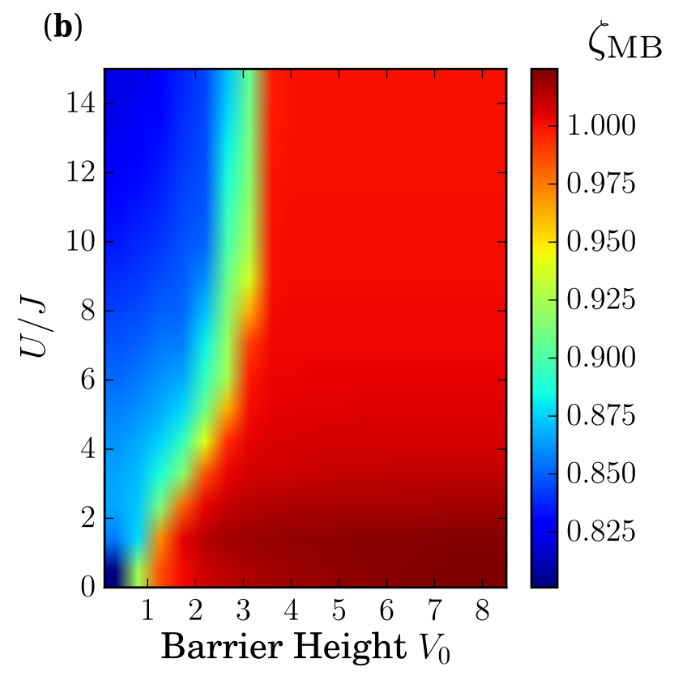}
\caption[Dynamic regimes predicted by initial states.]{\label{fig:doublePhaseDiag}\textit{Dynamic regimes predicted by initial states.} Using the ratio of the interaction to the tunneling energy, we can predict dynamics using information from static initial states. For a system of 15 lattice sites and 7 total particles, (\textbf{a}) the mean-field energy ratio depicts the Josephson regime in blue, the Fock regime in red for weak interactions, and the orange region does not reach the critical point $\zeta = 1$: the mean-field ratio fails to predict the symmetry-breaking phase transition for stronger interactions. In contrast, the many-body energy ratio (\textbf{b}) demonstrates a decisive Fock region in red where $\zeta_{\mathrm{MB}}=1$ for all interaction strengths.}
\end{figure}

The energy ratio $\zeta$ is one way of characterizing distinct dynamical regimes. The Josephson regime manifests in low-barrier cases, meaning the particles tunnel or oscillate back and forth between wells in a macroscopic manner. These are Rabi oscillations when $\zeta$ approaches $0$ . Above the single particle limit, for $\zeta < 1$, Josephson or plasma oscillations emerge. The critical point $\zeta = 1$ marks the phase transition; the high-barrier limit, or $\zeta > 1$, leads to macroscopic self-trapping of the condensate on one side of the well. The isolation of the junction, unlike that of superconducting circuits with external biases, allows for a more tenable parameter space to observe self-trapping \cite{malomed2014}. The ratio as a function of interaction strength or lattice depth $U/J$ and barrier height $V_0$ is represented in Figure \ref{fig:doublePhaseDiag} for a double well with 15 total sites and 7 total particles. The resolution in $U/J$ is 1, with $J=1$ and $U$ from 0 to 15. The resolution in $V_0$ is $0.5$. For small $U/J$, both $\zeta_{\mathrm{MB}}$ and $\zeta_{\mathrm{MF}}$ experience the $\mathbb{Z}_2$ critical region about $V_0=1$. For large interactions $U/J$, $\zeta_{\mathrm{MF}}$ in Figure \ref{fig:doublePhaseDiag}(\textbf{b}) does not reach the critical point and the mean-field measure thus fails to recognize the spontaneous symmetry breaking.

\begin{figure}
\centering
\includegraphics[width=0.47\textwidth]{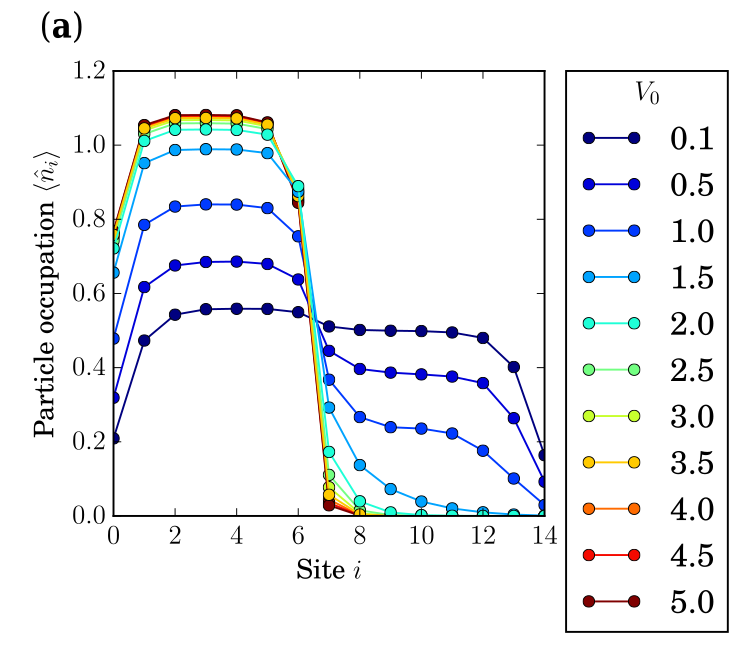}
\includegraphics[width=0.47\textwidth]{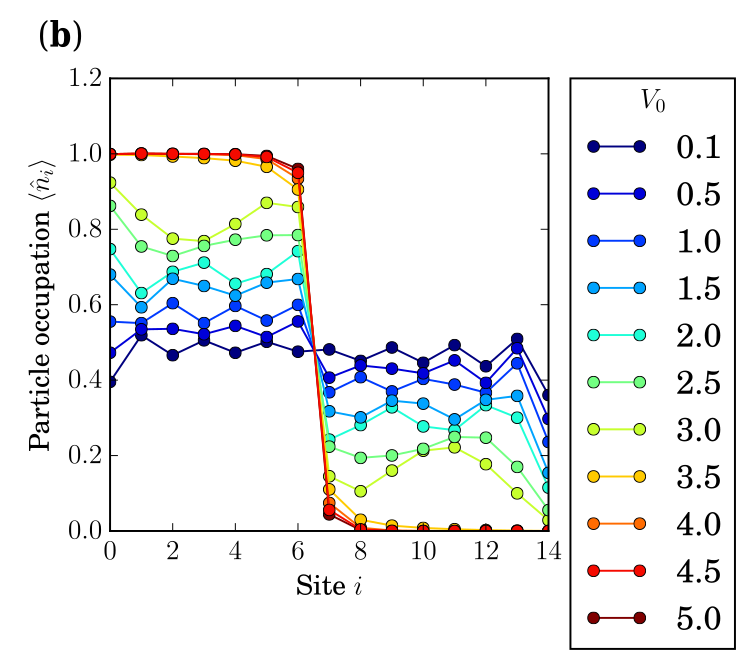}
\caption[Initial bound states through the Mott-superfluid critical point.]{\label{fig:initialsts}\textit{Initial bound states through the Mott-superfluid critical point.} State initialization traps the bosons largely in the left well for an optical lattice with 15 sites and 7 total particles. (\textbf{a}) Beyond the Mott-superfluid critical point, interactions are weak, $U=2$, $J=1$, and smooth superfluid behavior aligns more closely with mean-field theory. (\textbf{b}) In the Mott regime, interactions are strong, $U=15$, $J=1$, and low-lying excited modes inject deviations from mean-field theory that provide significant influence even with fluctuations as small as $\pm 0.05$ particles.}
\end{figure}

In Figure \ref{fig:initialsts}, the initial states immediately post-quench display a larger population imbalance $n_0$ toward the left well as the barrier height $V_0$ is increased. For weak interactions in Figure \ref{fig:initialsts}(\textbf{a}), the functions are smooth in portrayal of their condensed and superfluid nature. On the other hand, strong interactions instigate particle-hole pair formation as low-lying excited states of a Mott insulator, such as for $V_0=0.1,1$ in Figure \ref{fig:initialsts}(\textbf{b}), an effect which is not seen in mean-field simulations. The interaction of the particle-hole pairs with the condensate are a probable source of damping of the plasma oscillations in a closed system, as \cite{Levy2007} suggests, the interaction of the condensate with a non-condensed fragment is a source of phase decoherence -- an influence which may need consideration for high-precision applications such as superconducting quantum interference devices or BEC quantum simulators. For higher barriers such as $V_0=3$ in Figure \ref{fig:initialsts}(\textbf{b}), we find lowly-excited modes of the junction, with increasing excitations for higher barriers. In the dynamics, these modes are largely masked by stronger energy scales, but they are especially visible on the cusp of the transition to self-trapping, where the barrier is not yet high enough to completely trap the bosons in the left well, but the nonlinear disturbances from macroscopic tunneling are suppressed. The strongly-interacting Fock regime is reached for larger barrier heights than for the weakly-interacting regime due to repulsive interactions between atoms in the left well.

\subsection{Dynamic regimes and Fock measures}
\label{sec:dynamicFock}

The three dynamic regimes previously identified in isolated bosonic Josephson junctions are Rabi, Josephson, and Fock regimes \cite{albiez2005}. For weakly-interacting systems, where $U$ is small, these regimes are well-characterized with mean-field theory. In order to see the effects of a Mott insulator on the dynamics, we must have a number of particles that is approximately commensurate with the number of sites, though the presence of the barrier influences the Mott behavior \cite{batrouni2002}. Due to the nature of our state initialization, choosing a number of particles that is about half the number of sites gives a filling factor close to one for the single-well initial potential. This unit filling becomes more important as we increase interaction strength. For example, in Figure \ref{fig:initialsts}, there are a total of 15 sites with a 1-site barrier, and 7 bosons in one well enables close to unit filling for larger barriers.

\begin{figure}
\centering
\includegraphics[width=0.95\textwidth]{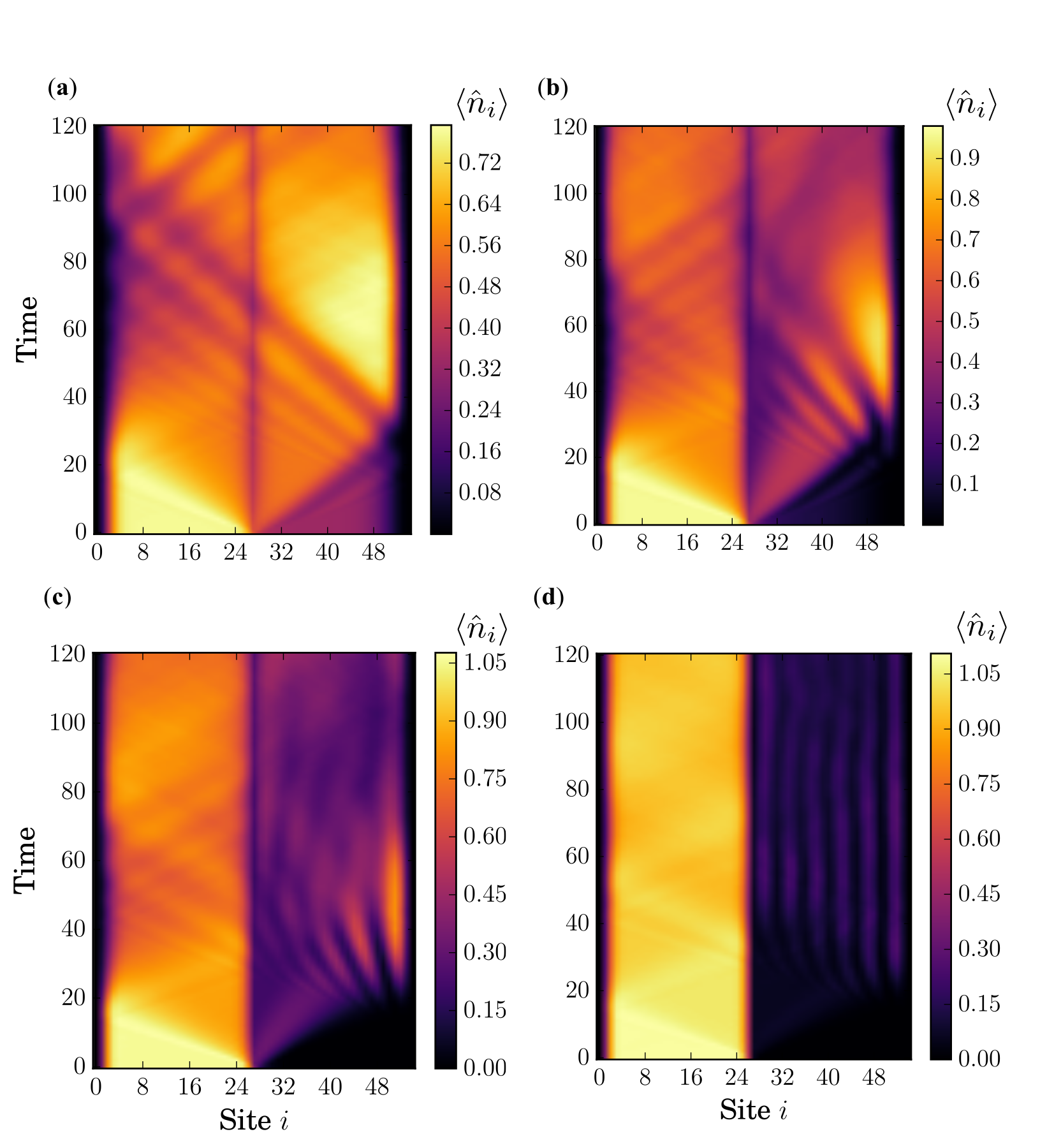}
\caption[Weakly-interacting spontaneous symmetry breaking transition.]{\label{fig:weakpt}\textit{Weakly-interacting spontaneous symmetry breaking transition.} Tunneling dynamics range from (\textbf{a}) the Josephson regime to (\textbf{d}) the Fock self-trapping regime for a system of 55 sites and 27 atoms, $U=0.3$ and $J=1$. The barrier height increases from (\textbf{a}) $V_0=0.2$ to (\textbf{d}) $V_0=2$. In the critical regions, (\textbf{b}) $V_0=0.4$ the dynamics are Josephson-like with interference patterns due to the interferometer nature of the double well. The diffraction fringes become more pronounced in the right well as the barrier height increases to (\textbf{c}) $V_0=0.6$ and becomes self-trapped in (\textbf{d}).} 
\end{figure}

Fock states or local particle numbers are one type of convenient measure of ultracold atom experiments. For realistic comparison with experiment, we characterize local particle densities as a function of time. Figure \ref{fig:weakpt} epitomizes the real time dynamics of a system with 55 lattice sites, 27 sites in each well, and 27 total particles, through the $\mathbb{Z}_2$ phase transition from Josephson in Figure \ref{fig:weakpt}(\textbf{a}) to Fock in Figure \ref{fig:weakpt}(\textbf{d}). The units are scaled by the tunneling parameter, where $J=1$, the units of time are $\hbar/J$, and typical lattice separation in experiments is $\simeq 1/2$ micron, with the units chosen such that the lattice constant $\Delta x = 1$ \cite{Cristiani2002,parsonsScience2016}. In this case, the interaction parameter $U=0.3$ is small and the barrier height is increasing from (\textbf{a}) $V_0=0.2$, (\textbf{b}) $V_0=0.4$, (\textbf{c}) $V_0=0.6$, (\textbf{d}) $V_0=2$, where the critical initial population imbalance is exceeded in (\textbf{c}) and (\textbf{d}) and the condensate is self-trapped as determined by $\zeta$ -- both mean-field and many-body versions.

\begin{figure}
\centering
\includegraphics[width=0.95\textwidth]{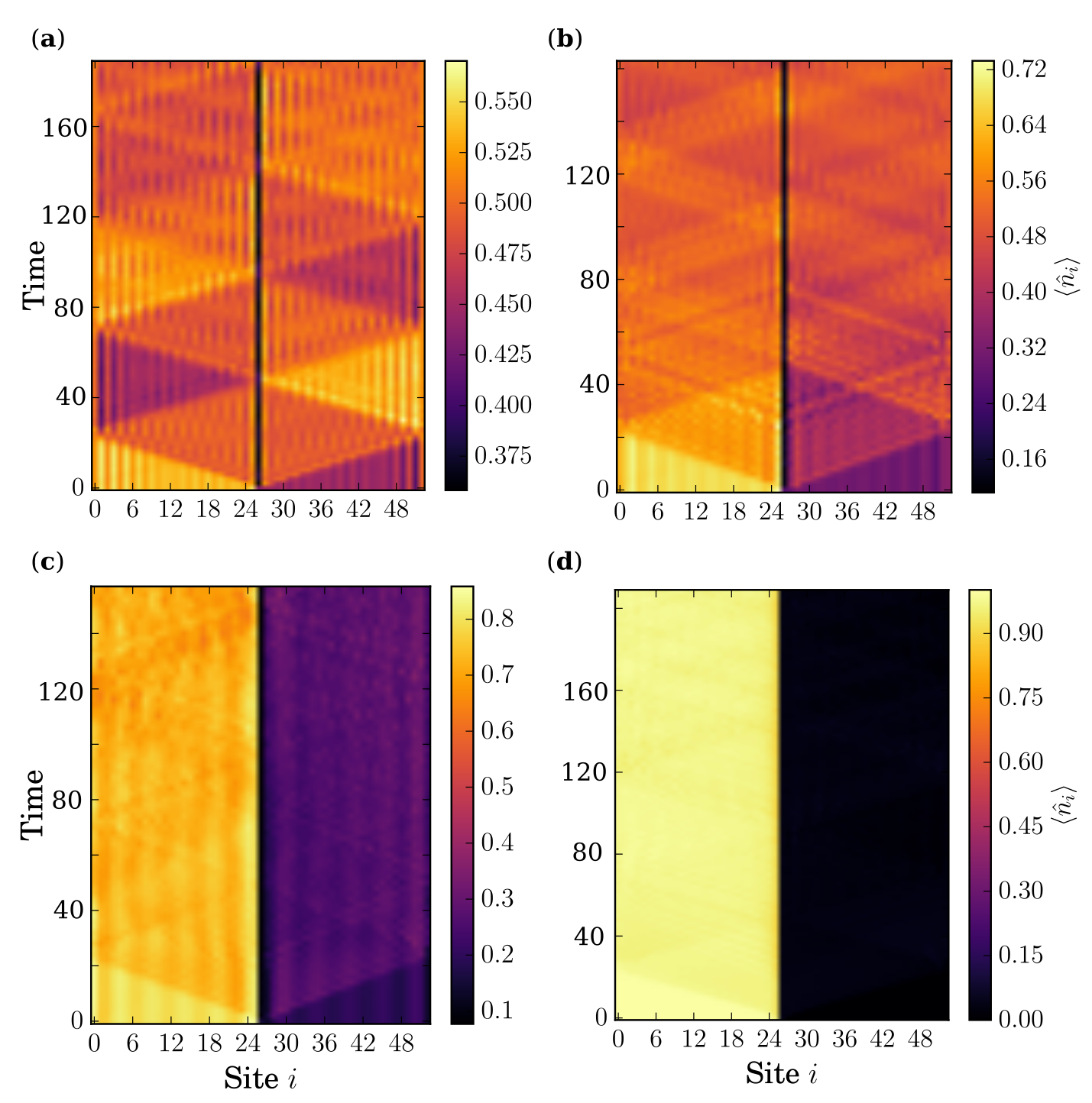}
\caption[Strongly-interacting spontaneous symmetry breaking transition.]{\label{fig:strongpt}\textit{Strongly-interacting spontaneous symmetry breaking transition.} For low barriers such as (\textbf{a}) $V_0=0.5$, particle-holes emerge across the entire lattice. Through the critical region of (\textbf{b}) $V_0=1$, the damping is large and phase coherence rapidly decays -- likely due to interaction of the quasiparticles with the condensate. Just beyond the critical point into the Fock regime for $V_0=3$ (\textbf{c}), the nonlinear waves that would otherwise dominate the dynamics are suppressed, making way for more timid modes of the double well, a result of the diabatic quench: these modes are sustained for unexpectedly long time scales. Finally, the bosons are self-trapped (\textbf{d}) for a larger barrier, $V_0=5$.}
\end{figure}

Next, we look at the transition from Josephson to Fock regimes when interactions are strong. In Figure \ref{fig:strongpt}(\textbf{a}), Mott behavior is dominant, and particle-hole pairs form in response to the rapid quench of the potential. These quasiparticles form faster than matter could possibly propagate from the left to the right well, indicating the sudden approximation may explain the phenomenon, as excited modes of the Mott insulator are visible in the Josephson regime.

Furthermore, the Josephson oscillations, likely a superfluid fragment or skin \cite{batrouni2002}, slosh over the top of the insulating quasiparticles and the interaction of the two fragments leads to a damping of the Josephson tunneling in Figure \ref{fig:strongpt}. As demonstrated with the energy ratio $\zeta$, the Josephson regime extends farther into the barrier height parameter space when interactions are strong, $U/J>3$. This extension of the parameter space is a mean-field effect, seen in both $\zeta_{\mathrm{MB}}$ and $\zeta_{\mathrm{MF}}$ diagrams of Figure \ref{fig:doublePhaseDiag}.

\begin{figure}
\centering
\includegraphics[width=0.47\textwidth]{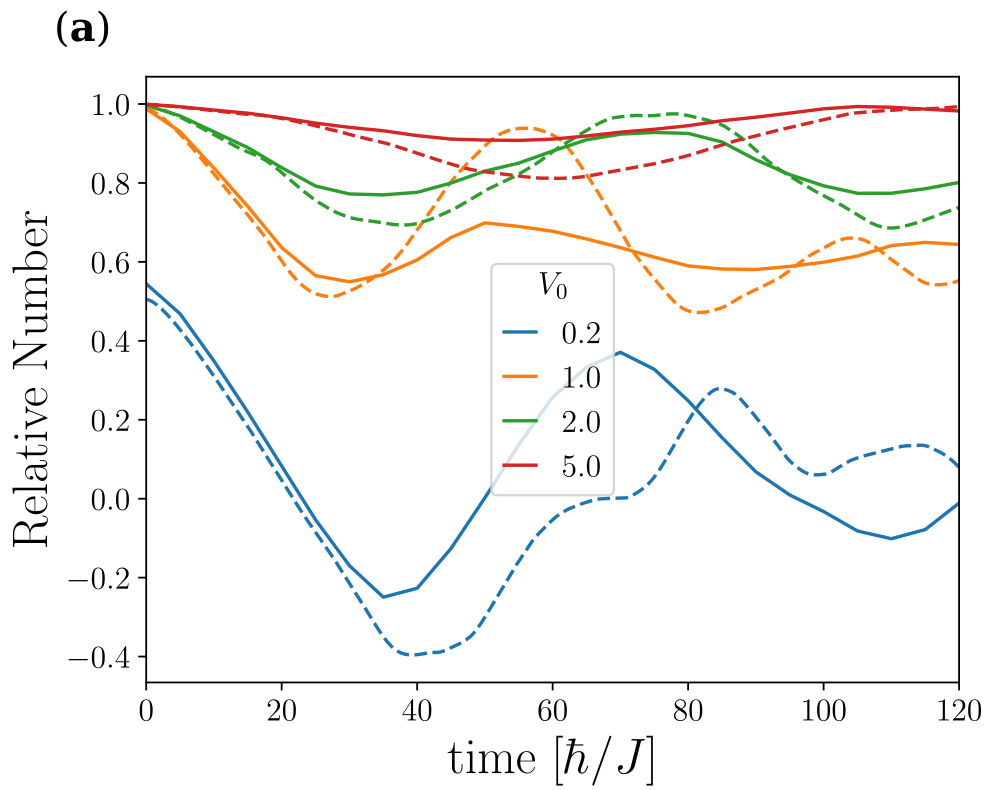}
\includegraphics[width=0.47\textwidth]{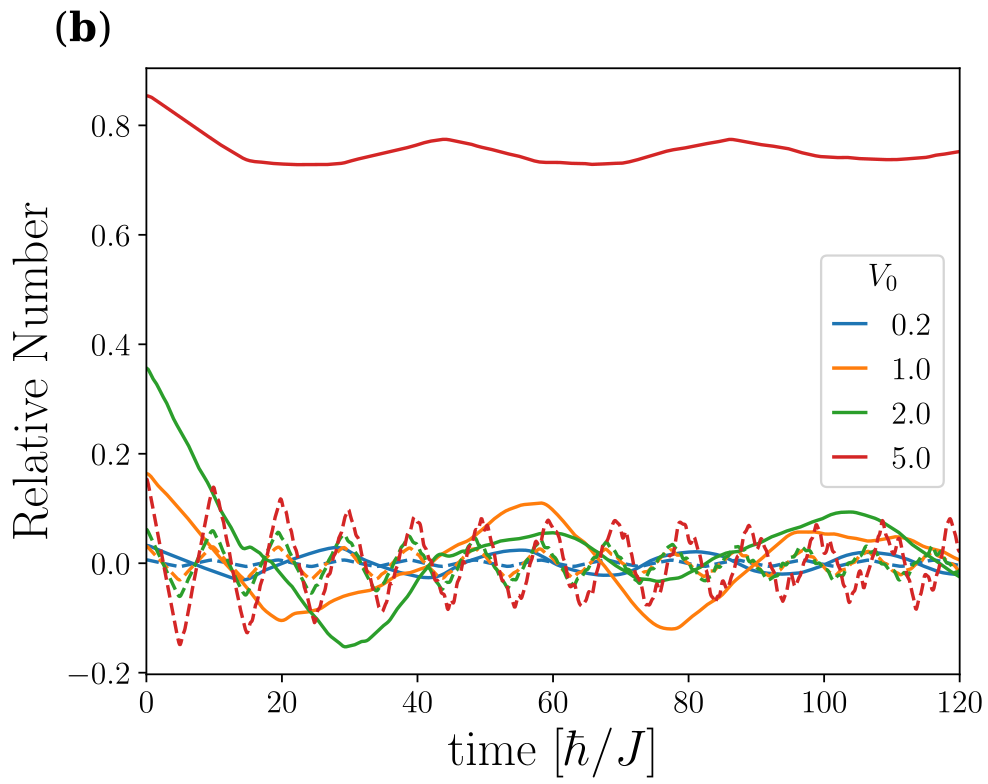}
\caption[Weak interaction similarities and strong interaction disparities: Time-evolving block decimation and the Gross-Pitaevskii equation.]{\label{fig:tebdgpe}\textit{Weak interaction similarities and strong interaction disparities: Time-evolving block decimation and the Gross-Pitaevskii equation.} For a system of 27 lattice sites and 14 particles, GPE (dashed curves) and TEBD (solid curves) simulations show (\textbf{a}) agreement of dynamical regimes for weak interactions and (\textbf{b}) a GPE failure to predict self-trapping for strong interactions. (\textbf{a}) Larger barrier heights $V_0=1,2,$ and $5$ are self-trapped for $J=1$ and $U=0.3$, and the low barrier $V_0=0.2$ is in the Josephson regime. (\textbf{b}) For $J=1$ and $U=30$, TEBD results (solid) predict self-trapping for a barrier height of $V_0=5$ whereas GPE results (dashed) predict Josephson oscillations for all barrier heights.}
\end{figure}

To further characterize the dynamics regimes, we consult the energy ratios suggested in Equations (\ref{eqn:zMF}) and (\ref{eqn:zMB}), which call for initial phase and number conditions. For simplicity, the initial relative phase is zero, leaving the initial relative number to determine the dynamic regime with this method. First we replicate research in \cite{albiez2005,Smerzi1997} supporting Josephson and Fock regimes for weakly-interacting $^{87}$Rb bosons from a mean-field perspective. Figure \ref{fig:tebdgpe}(\textbf{a}) reveals the qualitative agreement of mean-field GPE simulations and many-body TEBD for dynamics of the population imbalance or relative number for $J=1$ and $U=0.3$. The system consists of 27 total lattice sites, 13 sites in each well, and 14 particles, which enables approximate commensurate filling for one of the wells plus the barrier. For $V_0=0.2$, both mean-field and many-body simulations are in the Josephson regime as determined by $\zeta_{\mathrm{MB}}$ and $\zeta_{\mathrm{MF}}$. A quick method for diagnosing the dynamical regime is to plug the critical ratio $\zeta = 1$, and relative phase, $\phi_0\approx 0$, into Equations (\ref{eqn:zMB}) and (\ref{eqn:zMF}) and solve for the critical population imbalance $n_0$.  In this instance, the initial population imbalance $n_0 \approx 0.5$ is less than the critical value and thus it is in the Josephson regime. This conclusion supports the dynamical results: Josephson oscillations occur about the x-axis, so the average value of the function should be close to zero. This is akin to an average d.c. component of zero, meaning the particles are tunneling and moving freely across the barrier region. 

Additionally, for $V_0=1,2,$ and $5$, the initial population imbalance $n_0=1$ is greater than the critical value, in this case, for both many-body and mean-field simulations, suggesting that these systems are in the Fock regime. The next clue that particles are self-trapping behind the barrier, no longer able to freely tunnel, is the d.c. component of the relative number, as well as the smaller magnitude of the a.c. component. As pointed out in \cite{Smerzi1997}, there are sub-regimes within the macroscopic-self trapping region. The farther into the symmetry-breaking regime, the smaller the amplitude of the oscillations within the left well. Closer to the $\mathbb{Z}_2$ phase transition, the oscillations will resemble critical damping; the curves for $V_0=1$ (solid) demonstrate under-damping close to the critical damping region. Not coincidentally, $V_0\approx 1$ is where we find the $\mathbb{Z}_2$ critical region for small $U/J$ in the energy ratio $\zeta$ diagrams of Figure \ref{fig:doublePhaseDiag}.

We apply the same approach to Figure \ref{fig:tebdgpe}(\textbf{b}) using energy ratios. Both $\zeta_{\mathrm{MB}}$ and $\zeta_{\mathrm{MF}}$ predict Josephson dynamics for $V_0=0.2,1,$ and $5$, as evidenced by the small initial population imbalances for TEBD (solid) and GPE (dashed) simulations. The TEBD solid red curve depicts macroscopic self-trapping for $V_0=5$, which agrees with $\zeta_{\mathrm{MB}}$ in Figure \ref{fig:doublePhaseDiag}(\textbf{b}) for strong interactions. However, the GPE dashed red curve at the same barrier height falsely depicts Josephson oscillations with larger magnitudes -- a result that corroborates $\zeta_{\mathrm{MF}} <1$ for all barrier heights in Figure \ref{fig:doublePhaseDiag}(\textbf{a}), meaning there is no symmetry breaking transition. The disparity of MF from MB theory is not surprising, as often MF theories break down for strongly-interacting systems.

\begin{figure}
\centering
\includegraphics[width=0.45\textwidth]{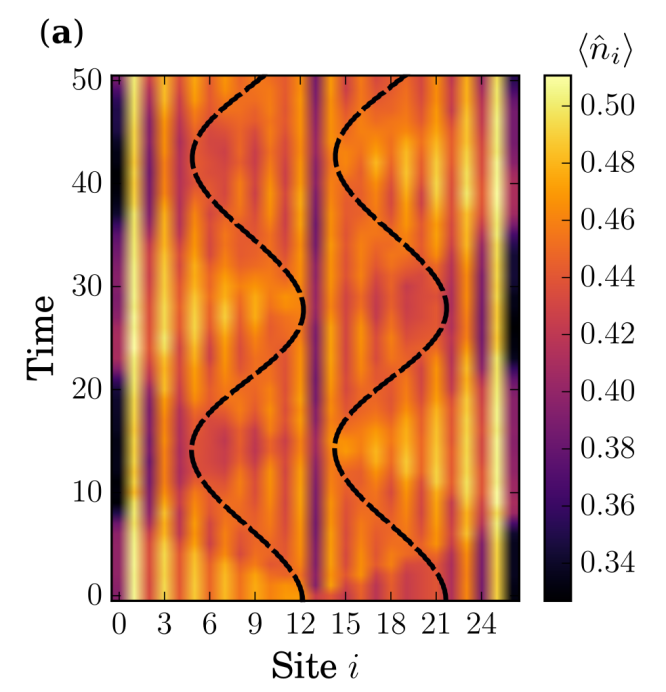}
\includegraphics[width=0.45\textwidth]{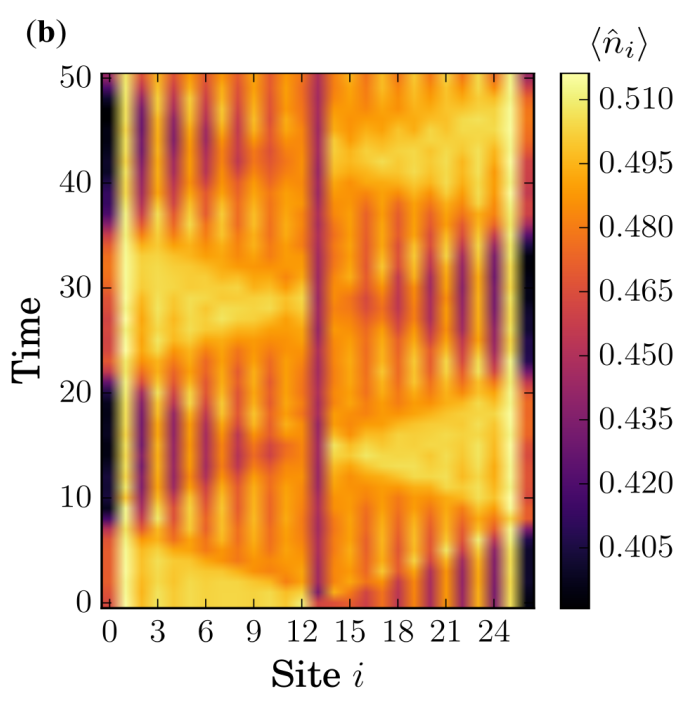}
\includegraphics[width=0.45\textwidth]{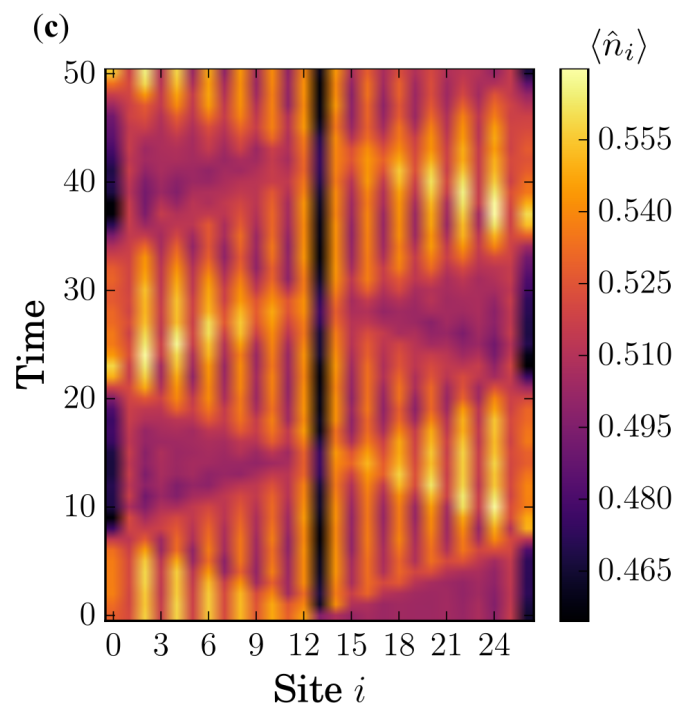}
\includegraphics[width=0.45\textwidth]{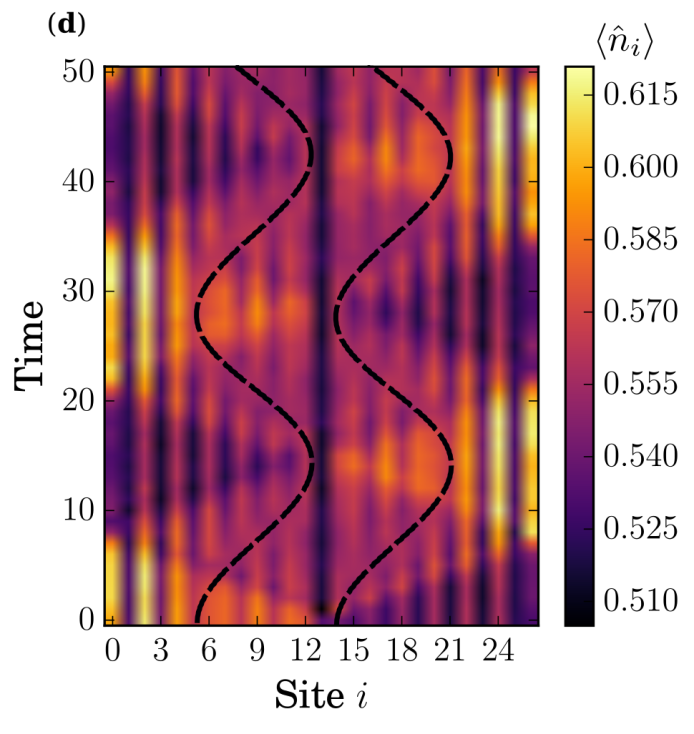}
\caption[Emergent solitons counter the double well symmetry.]{\label{fig:solitons}\textit{Emergent solitons counter the double well symmetry.} In the Josephson regime, (\textbf{a}) and (\textbf{d}) show pairs of propagating solitons -- one in each well -- potentially manifesting with a dressing of particle-hole pairs. In the plots, they appear as blurred sinusoids with a color halfway between the bright particles and the dark holes. The 27-site system has $V_0=0.2$, with $J=1$, $U=30$. The filling factor is not precise due to the presence of the barrier, so (\textbf{b}) 13 and (\textbf{c}) 14 particles are close to unit filling and do not exhibit soliton formation. Slightly (\textbf{d}) $N=15$ above or (\textbf{a}) $N=12$ below commensurate particle numbers seem to provide a ripe environment for these solitons, which oscillate in unison and against the mirror symmetry of the junction.}
\end{figure}

The final figure of our Fock space focuses on what appears to be soliton formation. Specifically, in the Josephson regime with strong repulsive interactions, particle-hole pairs form as excitations of a Mott insulator, seen as the bright and dark vertical bands in Figure \ref{fig:solitons}. The very dark band in the center is a density minimum due to the barrier. For $V_0=0.2$, with $J=1$, $U=30$, the double well has 27 sites, and the number of particles for each plot is (\textbf{a}) 12, (\textbf{b}) 13, (\textbf{c}) 14, and (\textbf{d}) 15. The magnitude of the quasiparticles, or the number separation between bright and dark bands, is larger closer to the edge walls compared with the quasiparticle magnitude in the center of the wells; we attribute this outcome to the counteraction of the double well boundaries, which asserts an effect akin to compressing an accordion. The smaller particle numbers, such as (\textbf{b}) and (\textbf{c}), display greater repulsion from the edge boundaries on sites $0$ and $54$. The larger particle numbers show the opposite effect and collide more directly with the walls due to the larger superfluid fragment sloshing between the wells. The solitons form for particle numbers immediately above or below commensurate filling. In this case, (\textbf{a}) has 12 particles and (\textbf{d}) has 15 particles, and the 13 and 14 particle cases are both relatively commensurate because of the thin barrier. The commensurate-filling plots do not demonstrate any soliton behavior, though the 14-particle case, perhaps closest to unit filling, illustrates the superposition of the superfluid fragment and the particle-holes when the quasiparticles disappear for brief regions in between macroscopic sloshing. The solitons in (\textbf{a}), beginning from site 12 in the left well and 21 in the right well, are $180$ degrees out of phase from the solitons in (\textbf{d}) beginning from sites 5 and 14. Both plots show the same soliton oscillation magnitude of about 9 sites, indicating there may be number symmetry about commensurate filling. The antisymmetric nature of the solitons on opposing sides of the barrier, as they propagate in parallel, provides a striking contrast to the double well symmetry. 

We identify these excitations as solitons because (i) they move at a velocity of about half the sound speed, (ii) they survive multiple collisions with the barrier as well as motion in the effective potential created by the accordion-like compression of the particle-hole background, and (iii) they occur only for particle numbers incommensurate with the number of lattice sites in the well.  The point of (iii) is that the extra particles above the Mott insulating background create a superfluid film or skin in the ``wedding cake'' structure well-known from other studies of the Bose-Hubbard model \cite{batrouni2002}. There is also some precedent for solitons in the particle-hole limit in a related context to the one here \cite{reinhardt2011}.  Finally we point out that in simulations doubling the number of particles and the number of sites we saw the same kind of solitons form, showing that this effect depends on the density, not the number of particles.

\subsection{Dynamic regimes and correlation measures}
\label{sec:dynamicCorrelations}

Beyond local particle numbers, we also examine correlation measures that highlight some advantages of matrix product state methods. For example, one method of identifying the superfluid-Mott phase transition is by tracking the depletion of the condensate, $D = 1 - \lambda_1/ (\sum_i \lambda_i)$, where the sum is over the $L$ eigenvalues of the SPDM, $\lambda_1 > \lambda_2 > \ldots > \lambda_L$, and taking the maximum $| \delta D |$ over a $J/U$ parameter quench as the critical point. The depletion gauges the amount of quantum gas living outside the BEC ground mode: $\lambda_1$ is this BEC mode, or the dominant eigenvalue of the SPDM, and the sum is over all eigenvalues of the single particle density matrix. The less depleted the condensate, the more superfluid the behavior will be. We emphasize that mean-field calculations cannot account for this many-body effect, with $D=0$ always.

\begin{figure}
\centering
\includegraphics[width=0.99\linewidth]{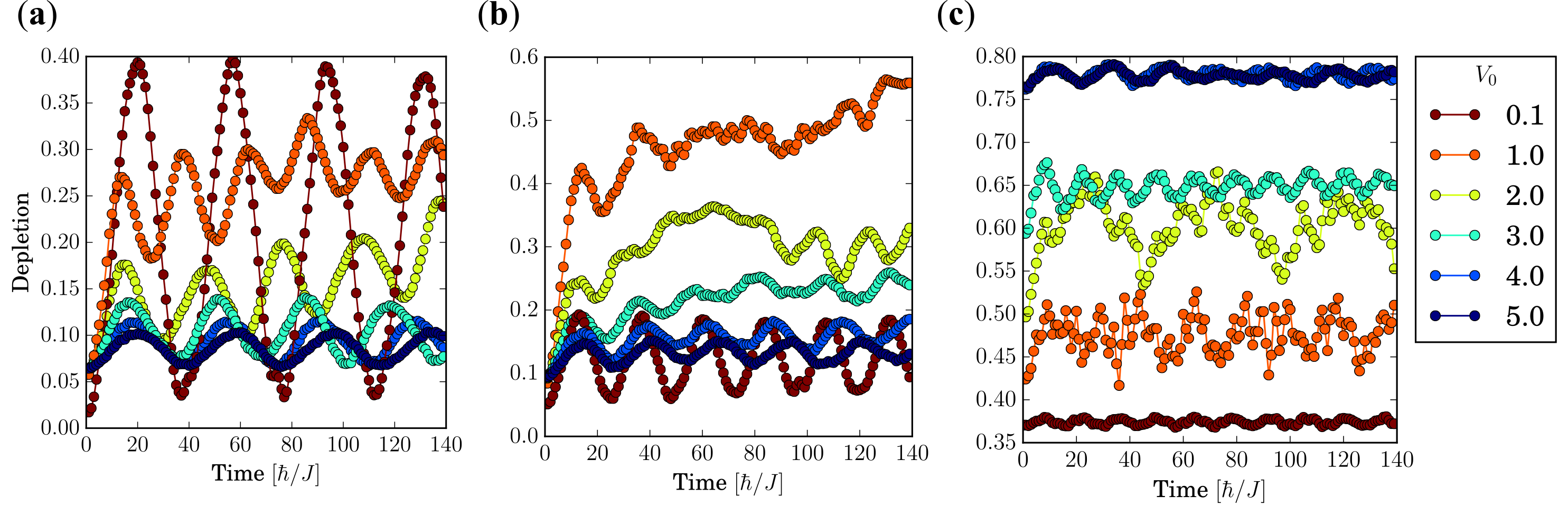}
\caption[Condensate depletion dynamics.]{\label{fig:depletion}\textit{Condensate depletion dynamics.} We consider a double well of 15 lattice sites and 7 atoms, and (\textbf{a}) the non-interacting case $J=1$ and $U=0$ most closely resembles Rabi oscillations, (\textbf{b}) the weakly-interacting case $U=2$ passes from Josephson to self-trapping at $V_0=1$, and (\textbf{c}) the strongly-interacting case $U=15$ is self-trapped for $V_0 \geq 4$. While the dynamics do not necessarily maintain the Mott or superfluid integrity they start with, (\textbf{c}) the Mott gap appears to influence the nearly fixed amount of the depletion when interactions are strong, as compared to the superfluid case.}
\end{figure}

We now consider the depletion of the left well for the same system size as in Figure \ref{fig:doublePhaseDiag}, 15 sites and 7 particles. In Figure \ref{fig:depletion}(\textbf{a}), we first address a system of non-interacting atoms, $U=0$, and the oscillations approach the Rabi limit for small barriers. The energy ratio $\zeta$ confirms that $V_0=2,3,4,$ and $5$ are all self-trapped. The curve for $V_0=1$ displays a dynamic empirical signature of the symmetry-breaking transition: the dynamics exhibit critical growth as the condensate depletes over time without returning to its initial condition. This divergence of the depletion from its initial condition is also seen in Figure \ref{fig:depletion}(\textbf{b}) near the critical point $V_0=1$, and this divergence diminishes the farther the barrier height is from its critical value.

In addition, Figures \ref{fig:depletion}(\textbf{a}) and (\textbf{b}) are uniformly initialized in the superfluid regime, but as the dynamics progress, higher barriers induce either oscillations or decays into more depleted states. In such cases, the systems are no longer strictly superfluid, nor are they in a Mott state. However, this initial state does influence the dynamics. For example, the initial depletion for superfluid states is continuous, while for the initial Mott states of Figure \ref{fig:depletion}(\textbf{c}),  the initial depletion is quantized, which is a signature of strong interatomic interactions. The energy ratio $\zeta_{\mathrm{MB}}$ confirms the highly-depleted states of $V_0=4$ and $5$ are self-trapped, and even the Josephson $V_0=0.1$ dynamics are substantially depleted due to the Mott influence.

The remaining figures demonstrate second order correlations via the $g^{(2)}$ measure, which we choose due to its accessibility in experiments \cite{Song2010}. The $g^{(2)}$ correlators quantify the fluctuation of particles for lattice sites $i$ and $j$ such that $g^{(2)}_{ij} = \langle \hat{n}_i \hat{n}_j \rangle - \langle \hat{n}_i \rangle \langle \hat{n}_j \rangle$. A positive $g^{(2)}$ means that the expectation value of measuring two particles simultaneously at sites $i$ and $j$ is larger than that of measuring the individual particles sequentially and vice versa. All $g^{(2)}$ plots have a minimum resolution of a single lattice site. These fluctuations are an important measure for determining limits of mean-field theory and provide insight into the overlapping nature of the two phase transitions. We use the standard normalization for optical lattices as opposed to quantum optics.  In lattices because only a small number of on-site number states are allowed, and the average occupation $\langle \hat{n}_j \rangle$ can be very small in places, one can get near-divergences in the usual normalized quantum optics $g^{(2)}$. A detailed discussion of this choice can be found in \cite{mishmash2008}.

\begin{figure}
\centering
\includegraphics[width=0.9\textwidth]{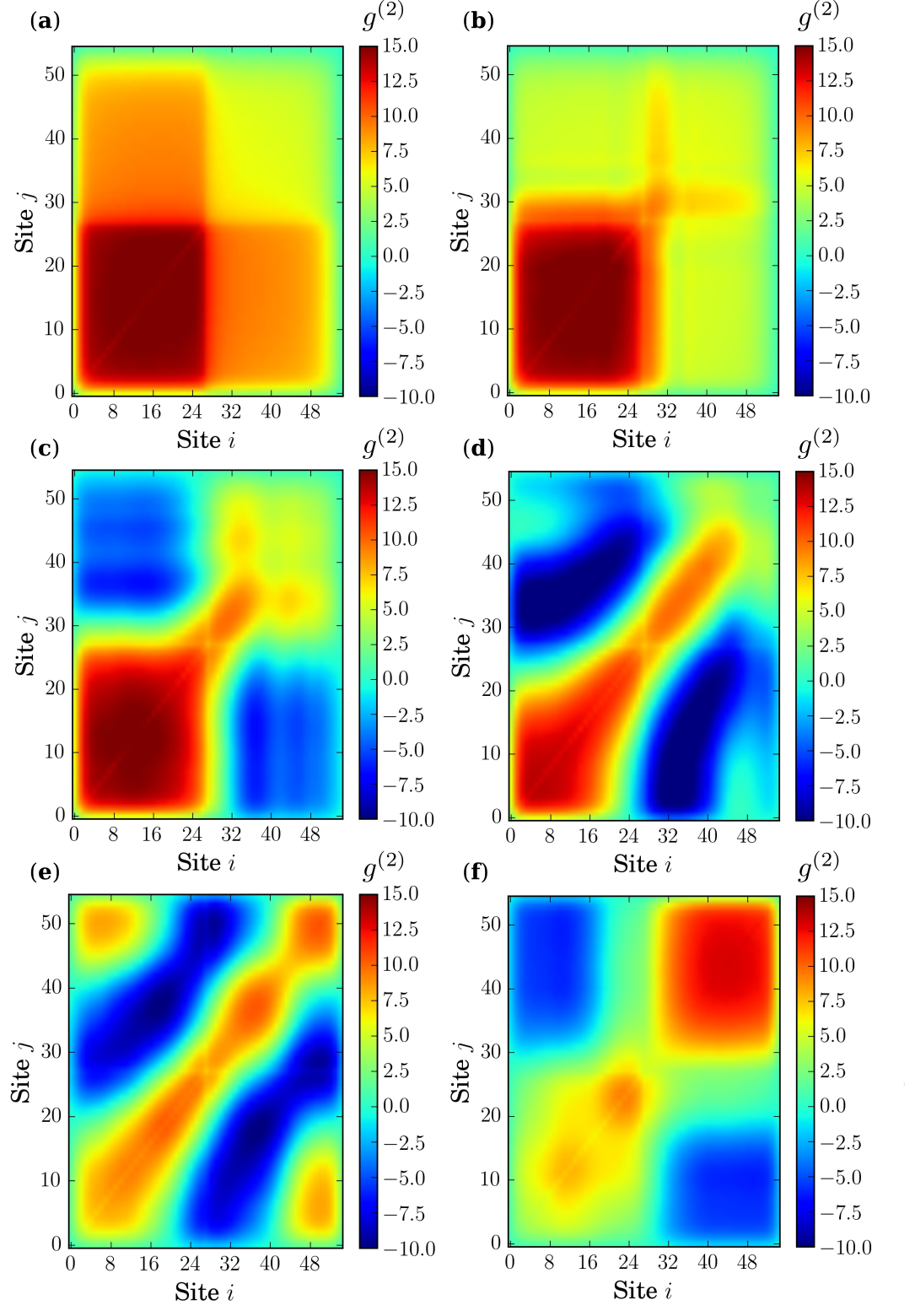}
\caption[Superfluid number fluctuation snapshots in the Josephson regime.]{\label{fig:g2weakJos} \textit{Superfluid number fluctuation snapshots in the Josephson regime.} For a junction with 55 sites and 27 particles, a barrier height of 0.2 and weakly-repulsive interactions, $J=1$ and $U=0.3$, number fluctuation dynamics convey superfluid Josephson transmission from (\textbf{a}) time $t=1$ in the left well to (\textbf{f}) $t=60$ in the right. At (\textbf{b}) time $t=5$ and (\textbf{c}) $t=10$, interference patterns develop from reflections off the barrier and at (\textbf{d}) $t=20$ the strong on-site or positive fluctuations begin to pull the off-site negative fluctuations away from the edges and toward the barrier. This creates the striped pattern (\textbf{e}) at $t=40$. (\textbf{f}) The BEC at $t=60$ predominantly occupies the right well. The large magnitude of the correlations is due to large superfluid occupation and thus long-range phase coherence.}
\end{figure} 

Figure \ref{fig:g2weakJos} delineates half of a period of Josephson oscillation in the weakly-interacting regime for a larger system size with 55 lattice sites, 27 particles, $V_0=0.2$, $J=1$ and $U=0.3$; in the first panel (\textbf{a}) at time $t=1$, fluctuations are, unsurprisingly, strongest in the left well where the condensate is initialized and decrease smoothly into the right well prior to collision with the far right wall. After this reflection, (\textbf{b}) a diffraction pattern emerges at $t=5$ for fluctuations $g^{(2)}\approx 4-9$. The reflected fluctuations (\textbf{c}) at $t=10$ create interference patterns both for the more highly-entangled regions within the two wells when $g^{(2)}>0$, and the non-entangled correlations between the two wells when $g^{(2)}<0$. (\textbf{d}) As the BEC begins to macroscopically tunnel to the right well at $t=20$, the negative fluctuations of the off-diagonals deepen, while the off-diagonals near the far left and right walls approach zero. (\textbf{e}) This trend continues for $t=40$ as the BEC collides with the right wall: the contrast of the off-diagonal pattern deepens from $g^{(2)}=-10$ to $6$. These positive fluctuations for correlations between sites $\sim 42-54$ and $\sim 0-18$ are likely due to momentary accumulation along the outer walls. Finally, (\textbf{f}) for $t=60$, the BEC largely occupies the right well, completing a half-period of the Josephson oscillation.


\begin{figure}
\centering
\includegraphics[width=0.9\textwidth]{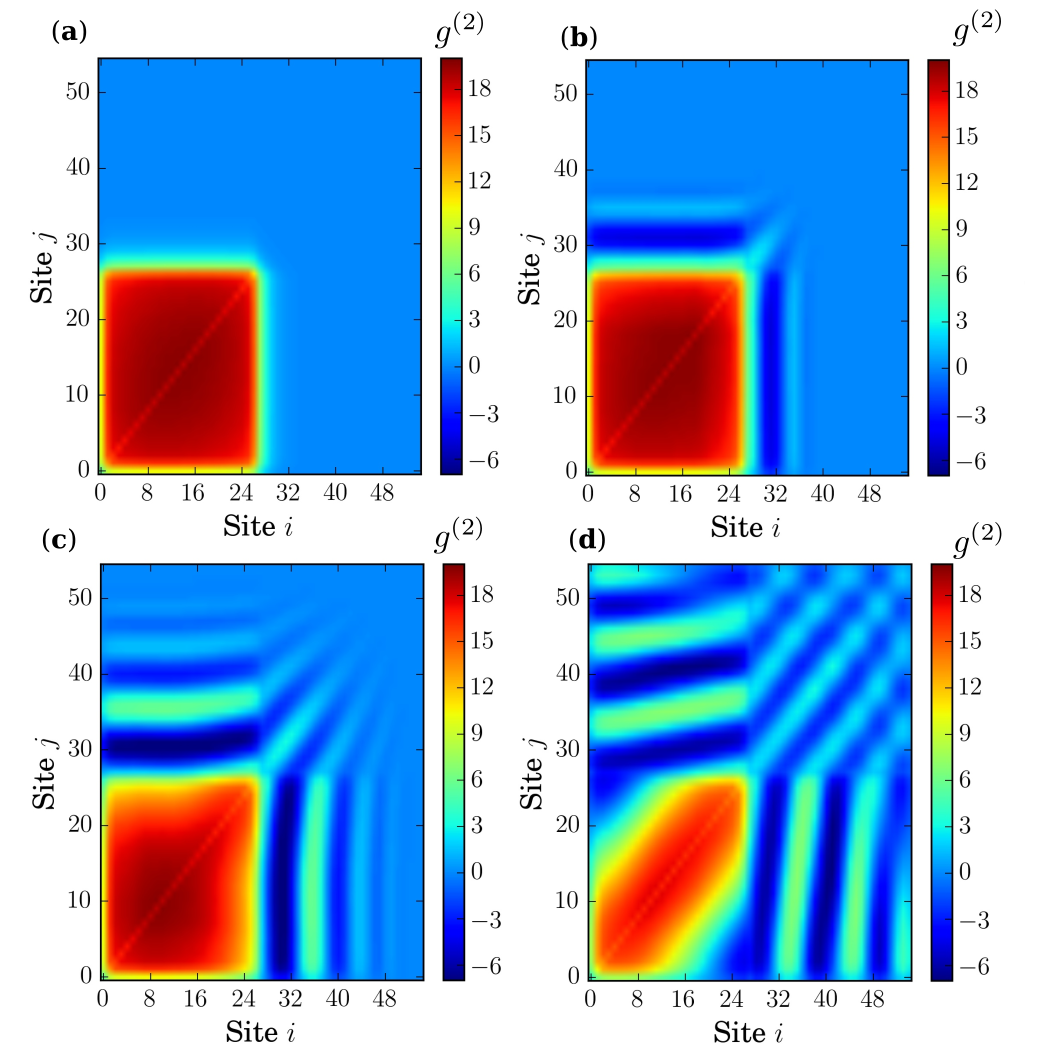}
\caption[Self-trapping fluctuation diffraction.]{\label{fig:g2weaktrap}\textit{Self-trapping fluctuation diffraction.} Interference patterns arise as the BEC slowly begins to escape the confines of the left well. $J=1.0$ and $U=0.3$ for a double well with 55 sites and 27 particles and barrier height $V_0=1.0$. As time evolves, (\textbf{a}) $t=1$, (\textbf{b}) $t=5$, (\textbf{c}) $t=10$, (\textbf{d}) $t=20$, the diffraction contrast deepens, but the macrosopic tunneling escape rate is much slower than the time scale of the experiment. Again, as the bosons are weakly-interacting, the long-range coherence leads to an increase in positive correlation magnitude.}
\end{figure}

The next series of $g^{(2)}$ plots in Figure \ref{fig:g2weaktrap} portray the time evolution for a self-trapped BEC as the transmitted fluctuations interact with the barrier, which is at site 26. The system size, particle number, and $J$ and $U$ values are the same as in the previous Figure \ref{fig:g2weakJos}, the difference being the barrier has been raised to $1$ and the condensate is in the Fock regime. The BEC is initially trapped in the left well (\textbf{a}) at $t=1$; then, transmission of the fluctuations through the barrier region produces diffraction patterns seen (\textbf{b}) at $t=5$ and (\textbf{c}) at $t=10$. The final panel (\textbf{d}) $t=20$ reveals an additional interference and more intense fluctuations from the rebound off the far right wall.

\begin{figure}
\centering
\includegraphics[width=0.9\textwidth]{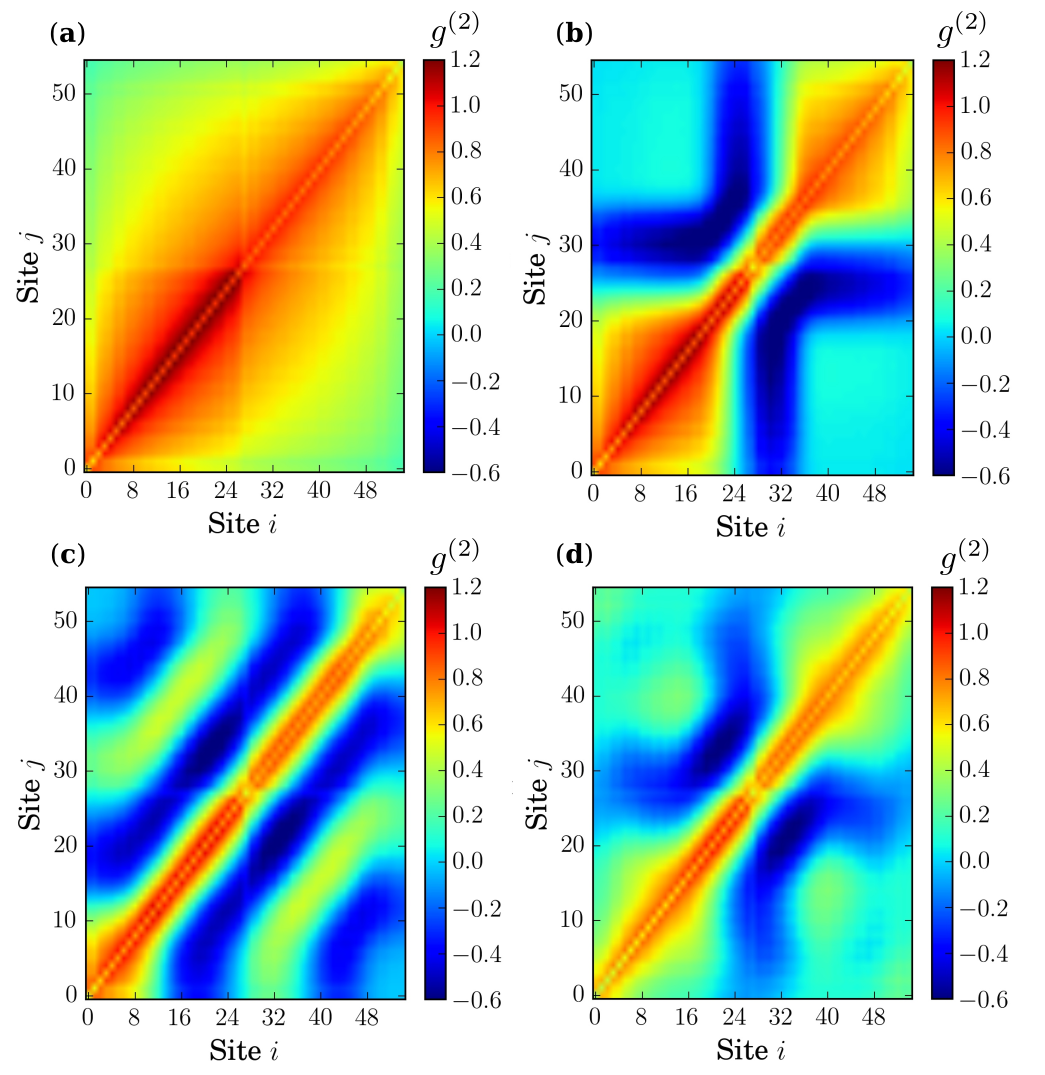}
\caption[Symmetric fluctuations near the self-trapping critical point.]{\label{fig:g2strongcrit}\textit{Symmetric fluctuations near the self-trapping critical point.} With strong interactions and barely into the Fock regime, $g^{(2)}$ number fluctuations spread from the left to right well before reaching a meta-stable equilibrium, where the on-site fluctuations are approximately even in both wells. $J=1$ and $U=30$ for a junction with 55 sites and 27 particles and barrier height $V_0=1$. (\textbf{a}) $t=1$, (\textbf{b}) $t=5$, (\textbf{c}) $t=10$, (\textbf{d}) $t=20$. }
\end{figure} 

Moving to strongly-interacting systems, Figure \ref{fig:g2strongcrit} illustrates a dissemination of the initial concentration of close-neighbor fluctuations in the left well (\textbf{a}) at time $t=1$. The system is the same size of 55 lattice sites and 27 particles, with $J=1$, $U=30$, and $V_0=1$. By (\textbf{b}) $t=5$, the number fluctuations have congregated in the right well, slightly weaker than the left. Because the junction is near the $\mathbb{Z}_2$ critical point, the fluctuations begin to equilibrate between the two wells without macroscopic oscillations. The magnitude of $g^{(2)}$ in this regime is substantially lower than in the superfluid cases of Figures \ref{fig:g2weakJos} and \ref{fig:g2weaktrap}, meaning these fluctuations may be more difficult to measure in experiment. For (\textbf{c}) $t=10$ and (\textbf{d}) $t=20$, we compare the fluctuations with Figure \ref{fig:strongpt}(\textbf{b}) at the same time, as it is the same system. The extra fluctuation bands appear at $t=10$ time because of a collision of the superfluid fragment with the far left boundary, and a smaller superfluid fragment collides with the far right wall. Because the population imbalance still favors the left well and the system is in a Fock regime, the symmetry is broken, which makes the mirror symmetry of the number fluctuations more surprising. 

Finally, the \textit{Fock flashlight} develops highly-localized nearest-neighbor correlations in the left well and weaker, delocalized correlations in the right well. For a double well with $J=1$, $U=30$, and $V_0=5$, the result is (\textbf{a}) $t=1$ the hilt of the flashlight comprises the self-trapped Mott insulator and the bright spot in the right well is immediately present -- another possible effect of the diabatic quench. Panel (\textbf{b}) at $t=5$ and (\textbf{c}) at $t=10$ exposes a familiar diffraction phenomenon, where positive number fluctuations build within the right well and collide with the barrier. Also at $t=10$, the flashlight begins growing an off-diagonal crossguard of Mott-like bright and dark bands that propagate symmetrically from the barrier toward the walls at (\textbf{d}) $t=20$. For (\textbf{e}) $t=30$, they have reflected off the external potential walls and by (\textbf{f}) $t=60$ they have begun to dissipate except directly along the barrier and the walls, where they have accumulated. The junction is smaller, 27 sites and 14 particles, because the strong interactions and high barrier required to observe this regime required a large bond dimension in TEBD, as we determined from convergence studies, so the larger system size was computationally intractable.

\begin{figure}
\centering
\includegraphics[width=0.9\textwidth]{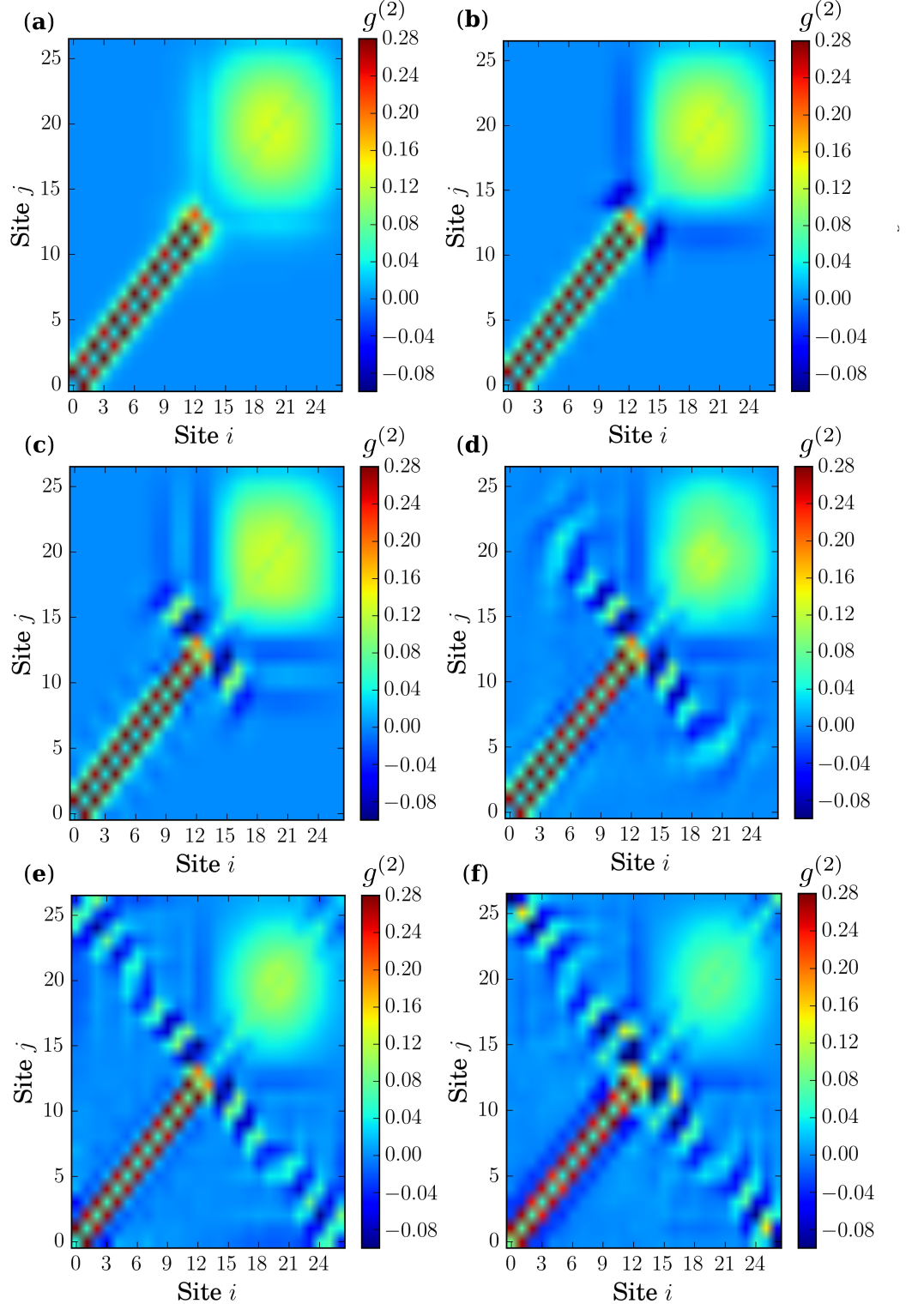}
\caption[The Fock flashlight effect.]{\label{fig:g2strongtrap} \textit{The Fock flashlight effect.} The Fock flashlight forms for strongly-interacting, self-trapped bosonic Josephson junctions. In this case we use a smaller system size of 14 particles on 27 sites due to larger computational demands for convergence. The barrier height is large $V_0=5$, with $J=1$, and $U=30$. (\textbf{a}) At time $t=1$, the fluctuations are highly localized in the left well and delocalized in the right well, creating the appearance of a flashlight. (\textbf{b}) The fluctuations begin to disperse by $t=5$, and for (\textbf{c}) $t=10$, (\textbf{d}) $t=20$, and (\textbf{e}) $t=30$, fluctuations propagate from the barrier outward as bright and dark bands on the anti-diagonal as fluctuations tunnel out of the right well and create vertical and horizontal diffraction bands. (\textbf{f}) Finally, by $t=60$, these flucutation patterns are dissipating save for the nearest-neighbors in the left well, where the bosons remain self-trapped. Strong interactions suppress long-range coherence, and thus the magnitude of the correlations is much smaller than in the weakly-interacting case.}
\end{figure} 

\section{Materials and Methods}
\label{sec:matmethods}

\subsection{Time-evolving block decimation (openTEBD)}
Figure \ref{fig:ITPRTP} depicts the method we use to simulate a reproducible initial state for double well dynamics. We first run imaginary time propagation using time-evolving block decimation (openTEBD) \cite{ref:openTEBD}; the right well of the potential is raised as shown in Figure \ref{fig:ITPRTP}(\textbf{a}). This provides an exponential decay in imaginary time rotations of the unitary operator that allows the high energy states to decay quickly, leaving us with the ground state of the potential largely occupying the left well. We then lower the right side of the double well at time $t=0$ as in Figure \ref{fig:ITPRTP}(\textbf{a}), so we no longer have a stationary state, and propagate forward in real time. We choose this protocol for similarities with experiment \cite{Cristiani2002} and for easier reproducibility.

The total error from the simulations can be written $\epsilon = \epsilon_{\mathrm{method}}+\epsilon_{\chi}$, where $\epsilon_{\mathrm{method}}$ is due to a combination of errors, the largest of which are the errors due to Trotter decomposition as well as that due to local dimension restrictions. $\epsilon_{\chi}$ stems from the Schmidt truncation, which is due to truncation of the Hilbert space.  From detailed convergence studies, we estimated the local dimension, d = max filling + 1, the bond dimension $\chi$, and the real and imaginary timesteps. We studied convergence of the dynamics as a function of local dimension, bond dimension, and time step. The low-barrier regime with weak interactions required small time steps, on the order of $10^{-4}$, and the high-barrier regime with strong interactions required large bond dimension, on the order of $\chi=100$ for the 27-site system. This slow convergence for the strongly-interacting Fock regime appears to indicate high entanglement that pushes the limits of matrix product state simulations.

The open source openTEBD software can be downloaded from sourceforge.net \cite{ref:openTEBD}, however, the authors strongly encourage consideration of open source matrix product state methods (OSMPS), also available on sourceforge.net \cite{ref:OSMPS} as a more up-to-date simulation tool \cite{wall2012,schollwoeck2005,Scholl2011}. 

\subsection{Gross-Pitaevskii equation (GPE)}
The Gross-Pitaevskii equation is a mean-field description of cold and dilute Bose gases. This description assumes a perfectly condensed state, ignoring fluctuations about the mean such that the depletion is always zero. A quasi-1D GPE is given by,
\begin{equation}
i \hbar \frac{\partial}{\partial t} \Psi(x,t) =\biggl[ \frac{-\hbar^{2}}{2m} \frac{\partial^{2}}{\partial x^{2}}  + V(x,t) + g_{\mathrm{1D}}|\Psi(x,t)|^{2} \biggr] \Psi(x,t). \label{eq:1dGPE}
\end{equation}
where tight harmonic confinement has been assumed in the transverse directions \cite{carretero-gonzalez_nonlinear_2008}.
The order parameter, $\Psi$, corresponds to the single-particle wave function. The nonlinear interaction parameter, $g_{\mathrm{1D}}$, is proportional to the $s$-wave scattering length, and results from assuming binary contact between atoms; note, from Table \ref{table:hamstable}, $g_{\mathrm{1D}} = g/2 \pi \ell_{\bot}^{2}$, with transverse harmonic oscillator length $\ell_{\bot} = \sqrt{\hbar/m \omega_{\bot}}$, where $\hbar$ is the reduced Plank constant, $m$ is the atomic mass, and $\omega_{\bot}$ is the transverse confining angular frequency.

Numerics are computed on a discretized version of Equation (\ref{eq:1dGPE}), called the Discrete Nonlinear Schrodigner Equation (DNLS),
\begin{equation}
i \hbar \frac{d}{dt}\psi_{i} = -J(\psi_{i+1} + \psi_{i-1}) + g_{\mathrm{1D}}| \psi_{i}|^{2} \psi_{i} +V_{i} \psi_{i},
\end{equation}
where $J$, $V_{i}$, and $g_{\mathrm{1D}}=U$ correspond to the values given in Table \ref{table:hamstable}. In the DNLS, $|\psi_{i}|^{2}$ corresponds to the expectation value of $ \langle \hat{n}_i \rangle$ on site $i$, we normalize to the number of particles $\sum_{i=1}^{L} |\psi_{i}|^{2} = N$, and $L$ is the number of lattice sites. The DNLS can also be derived from a mean-field approximation of the BHH, or a direct discretization of the GPE \cite{mishmash_ultracold_2009}.

Similar to many-body dynamics, an initial ground state is calculated using imaginary time propagation on a single potential well, Figure \ref{fig:ITPRTP}(\textbf{a}), and real time propagation with the barrier well dropped as in Figure \ref{fig:ITPRTP}(\textbf{b}). Imaginary time propagation is calculated using the fourth-order Runge-Kutta method, and real time propagation is performed with the LSODA implementation in SciPy \cite{scipy}, as originally described in \cite{petzold_automatic_1983}. This algorithm automatically chooses between the Adams-Bashforth method (an explicit numerical method) for non-stiff time evolution and backward differential formula (BDF) method (an implicit numerical method) for stiff evolution.
This scheme was used because the dynamics for larger $U$, $V_{0}$, and, $N$ occasionally require time-steps so small as to be computationally restrictive for explicit methods, in other words the problem behaves as a stiff differential equation. Because it is difficult to know when exactly the equations are stiff, the automatic selection scheme only uses the computationally more intensive BDF method when necessary.

\subsection{Sudden Approximation}
For strongly-interacting systems in the Josephson regime, particle-hole pairs form immediately as bright and dark bands. They form too quickly for atoms to propagate across the lattice, and they are instead due to the diabatic quench of the potential from a single to a double well. To support this hypothesis, we perform analytical sudden approximation calculations using exact diagonalization together with second-order perturbation theory. The calculations also support an analogous phenomenon that occurs in the Fock regime: the sudden projection of the single well ground state to the double well excites macroscopic modes of the double well. 

We perform the calculations on manageably small system sizes for both even and odd numbers of sites, one for 2 particles on 4 sites and the other 3 particles on 5 sites. We implement the sudden quench by projecting the Fock basis of 2 sites onto 4 sites and 3 sites onto 5 sites. The purpose is to demonstrate the immediate response of the initial state to such a quench. For example, for the 4-site case, we do the calculation three ways: (i) we use the ground state of the 2-site system for $J=0$, or an initial state $|\, 1\ 1\ 0\ 0\: \rangle$, (ii) we use a ground state that is a superposition of the possible states for the 2-site system, $\alpha_1 |\, 2\ 0\: \rangle + \alpha_2 |\, 1\ 1\: \rangle + \alpha_3 |\, 2\ 0\: \rangle$ for small, finite $J$; (iii) we use degenerate perturbation theory for pertubative tunneling, which agrees with the exact diagonalization of method (ii). In all cases, we take $J=1$ and $U=30$ so as to focus on the strongly interacting regime. 

\begin{figure}
\centering
\includegraphics[width=0.5\textwidth]{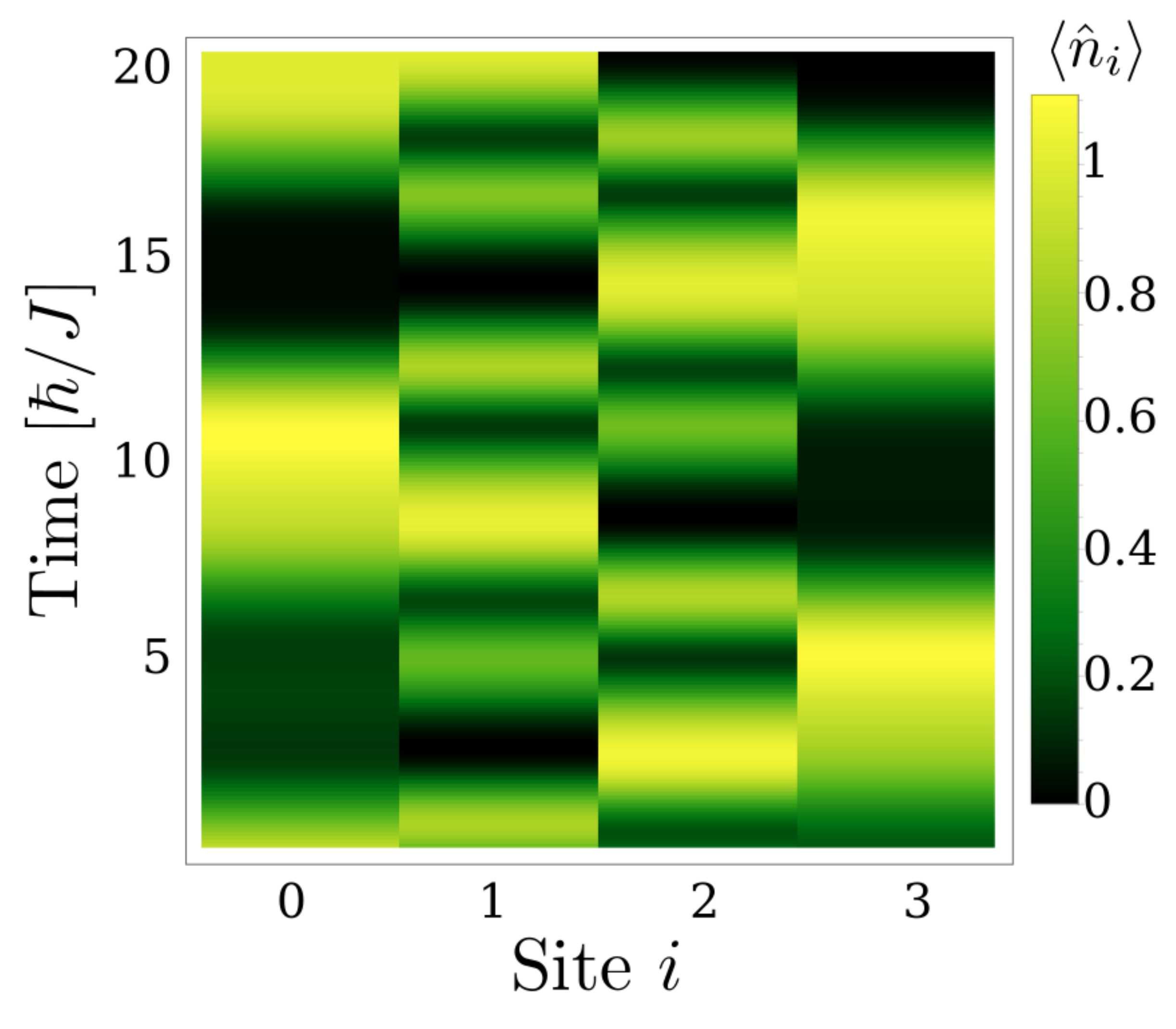}
\caption[Sudden approximation demonstrates propensity for excited mode formation.]{\label{fig:sudden}\textit{Sudden approximation demonstrates propensity for excited mode formation.} Analytical dynamics calculated with the sudden approximation and exact diagonalization on a system of 2 bosons on 2 sites projected onto 4 sites with $J=1$ and $U=30$. Unsurprisingly, the overall dynamics oscillate back and forth within the well. It is the more microscopic behavior, though, that yields insight into the accordion-like effects we see in TEBD simulations. The smaller peaks depict excited modes $|\, 1\ 0\ 1\ 0\: \rangle$ and $|\, 0\ 1\ 0\ 1\: \rangle$ that demonstrate a preference for mode formation with the same symmetry; this symmetry applies both on the optical lattice scale -- as particle-hole pairs -- and on the scale of the potential as a whole -- as excited modes of the double well.}
\end{figure}

In the 4-site case for method (ii), we first calculate the eigenvectors for the $L=4$, $N=2$ system and the ground state of the $L=2$, $N=2$ system using exact diagonalization, which provides a good initial state for the dynamics. Then for unitary time evolution, the only non-zero elements are from $L=4$ eigenvectors that have any superposition with the initial state, which in the new basis is a superposition of $|\, 2\ 0\ 0\ 0\: \rangle$, $|\, 1\ 1\ 0\ 0\: \rangle$, and $|\, 0\ 2\ 0\ 0\: \rangle$. We project the eigenvectors of the 10-dimensional $L=4$ Hilbert space onto the basis vectors to create number density vectors for each of the 4 lattice sites, which provides a common ground for comparison with TEBD simulations. The error in the calculations is due to the approximation itself, which assumes a perfect projection from the $L=2$ basis to the $L=4$ basis; this simple projection does provide insight into the diabatic quench in TEBD and its influence on exciting modes in the dynamics. These modes manifest as Fock states $|\, 1\ 0\ 1\ 0\: \rangle$ and $|\, 0\ 1\ 0\ 1\: \rangle$ and can be seen in Figure \ref{fig:sudden} as the smaller, nonlinear spikes between the larger oscillations. These Fock states, while by definition are not particle-hole pairs due to the limited particle number, support the propensity for the formation of local states of the same symmetries such as those in Figures \ref{fig:strongpt}(\textbf{a}) and (\textbf{b}) and Figure \ref{fig:solitons} -- i.e. particle-hole pairs. In method (iii), we calculate the eingenvectors instead with perturbation theory in the large $U$ limit and obtain qualitatively the same result. Method (i) was unsuccessful in demonstrating the dynamics because the initial state was purely the Mott ground state of the 2-site system $|\, 1\ 1\ 0\ 0\: \rangle$; this approximation is insufficient because the accordion-like excitations will only form for non-zero $J$.

The same methods were applied to the $L=5$ case. The odd number of sites case makes an important difference in the dynamical excitations as compared to the $L=4$ case: even without an explicit barrier, we see mode formation on two distinct sides of the single well. The Fock state $|\, 0\ 3\ 0\ 0\ 0\: \rangle$ is clearly an excited mode of the first 3 sites, with an even symmetry among these 3 sites. Although this state is energetically improbable, the overlap of the 3-site basis with the 5-site basis places a weight on the first 3 sites that over-emphasizes this projection compared with TEBD. However, the presence of an excited mode of one side of the potential supports the instanteous excited mode creation in Figure \ref{fig:strongpt}(\textbf{c}). The number density state $|\, 0\ 1\ 0\ 1\ 1\: \rangle$ further supports this conclusion, as the boson on the second site demonstrates the same symmetry as a ground mode for these first three sites, and the particles on sites 4 and 5 are Mott ground states of these last two sites; thus, the dynamics appear to manifest through excited modes of each side of the well. We further note the importance of symmetry when recognizing such manifestations of excited modes, as an even $L$ with the barrier breaks the pairing tendency and injects an extra level of excitation. The same argument holds for excited modes of the overall potential.

\section{Conclusions}
\label{sec:conclusion}
In conclusion, we have computationally demonstrated comparisons of mean-field and many-body dynamical regimes based on the interplay of the Mott--superfluid U(1) and the Josephson--self-trapping quantum $\mathbb{Z}_2$ phase transitions (QPTs) in a bosonic Josephson junction. Using initial state information, we design a many-body energy ratio to predict these dynamical regimes; and we find that mean-field theory breaks down as the repulsive interaction strength increases and fails to predict the spontaneous symmetry-breaking transition. Additionally, we identify a number of strictly many-body phenomena in these dynamical regimes that increase in importance for higher-precision applications: long-lived particle-hole pairs and low-lying double well modes arise due to a sudden quench of the potential and can act as signatures of QPTs - even in mesoscopic systems experimentally accessible on present quantum simulator platforms \cite{vanfrank2016}. Other many-body signatures include soliton formation in the Josephson regime and the $g^{(2)}$ Fock flashlight in the self-trapping regime: these effects manifest in quantities that are measurable experimentally in architectures ranging from cold atoms to nonlinear optics to superconductors. While our present study is discrete by nature of the optical lattice, future work can build on the results in the strongly interacting case by systematically increasing the number of lattice sites toward the continuum limit, for instance in a Tonks-Girardeau gas \cite{Zollner2008,Lode2009,Lode2012,bloch2012}. Further extensions of this research would incorporate fermions for a more encompassing unification with superconductors \cite{Valtolina2015,burchianti2018}. With the number of applications relying on two-mode models of the double well, the present study encourages investigation of cold atom double well experiments with optical lattices as highly-controllable quantum simulators of many-body effects in systems previously assumed to be mean-field.

\begin{acknowledgements}
The authors extend appreciation and gratitude toward Ver\'{o}nica Ahufinger and Anna Sanpera for inspiration and collaboration in the conception of these studies. Many heartfelt thanks also to Daniel Jaschke, Gavriil Shchedrin, and Meenakshi Singh for insightful discussions and new perspectives. Computations were performed using high performance computing resources at the Colorado School of Mines in conjunction with the Golden Energy Computing Organization. This material is based in part upon work supported by the US National Science Foundation under grant numbers PHY-1520915, PHY-1207881, PHY-1306638, OAC-1740130, as well as the US Air Force Office of Scientific Research grant number FA9550-14-1-0287. This work was also performed in part at the Aspen Center for Physics, which is supported by National Science Foundation grant PHY-1607611.
\end{acknowledgements}

\bibliographystyle{spmpsci} 



\end{document}